\documentclass[a4paper,11pt]{article}
\pdfoutput=1 
\usepackage{jcappub} 
\usepackage[T1]{fontenc} 
\usepackage{graphicx}  
\usepackage{todonotes}
\usepackage{changepage}
\usepackage{academicons}
\usepackage{natbib}
\usepackage{multirow}
\usepackage{orcidlink}
\usepackage{subcaption}
\usepackage{academicons}
\usepackage{soul}
\definecolor{orcidlogocol}{HTML}{A6CE39}

\usepackage{aas_macros}

\title{The Prospect from the Upcoming CMB Experiment LiteBIRD to Discover Axion-like Particles Using Milky Way}

\author{Harsh Mehta\orcidlink{0009-0007-4664-4820},}
\author{and Suvodip Mukherjee\orcidlink{0000-0002-3373-5236}}

\affiliation{Department of Astronomy and Astrophysics, Tata Institute of Fundamental Research, 

Homi Bhabha Road, Mumbai- 400005, India}

\emailAdd{harsh.mehta@tifr.res.in}
\emailAdd{suvodip@tifr.res.in}

\begin{document}
\abstract{
The existence of axion-like particles (ALPs) can be probed from their signatures in the Cosmic Microwave Background (CMB) due to the photon-ALP resonant conversion over the mass range of ALPs that matches with the effective mass of photons in the plasma in the astrophysical systems. Such a conversion can also occur in the Milky Way halo and disk and can cause a unique spatial and spectral distortion. The signal is highly non-Gaussian and cannot be measured precisely by the usual power-spectrum approach. We devise a new technique to search for this signal from the upcoming full-sky CMB experiment LiteBIRD using its multi-frequency band using a template-based spatial profile of the ALP distortion signal. This technique captures the large-scale non-Gaussian aspects of the ALP distortion signal in terms of a spatial template and makes it possible to search for any non-zero ALP signal. We show that the inference of the ALP coupling using the template-based technique from LiteBIRD can provide constraints on the coupling constant approximately $ g_{a\gamma} < 6.5 \times 10^{-12} \, \mathrm{GeV}^{-1}$ for ALP masses below $10^{-14}$ eV at 95\% confidence interval which is an order of magnitude better than the current bounds from CERN Axion Solar Telescope (CAST) at  $g_{a\gamma} < 6.6 \times 10^{-11} \, \mathrm{GeV}^{-1}$, This shows the capability of future multi-band CMB experiment LiteBIRD in opening the discovery space towards physics beyond the standard model.
}

\maketitle
\flushbottom

\section{Introduction}
The Cosmic Microwave Background (CMB) is the primordial radiation from the epoch of recombination, that follows a Planck black-body spectrum, with a monopole temperature of about 2.7255 K \cite{Dodelson:2003ft,Fixsen_1996,Fixsen_2009}. The CMB shows spatial fluctuations in temperature and polarization, which are referred to as anisotropies, which are of the order of $10^{-5}$ \cite{hanson2009estimators,Fixsen_1996,Fixsen_2009,Dodelson:2003ft}. These are generated either before the decoupling of photons from the primordial fluid (called primary anisotropies such as the Sachs-Wolfe effect, acoustic oscillations,  etc.), or after decoupling (called secondary anisotropies such as lensing, reionization, etc.) \cite{hu2002cosmic,hu1995toward,hu1997cmb,hu1998complete,lewis2004cmb,1972CoASP...4..173S,Smith_2007,lewis2006weak,adam2016planck,barkana2001beginning}.  In addition to these, the CMB is also expected to bear tiny deviations from the Planck black-body spectrum, due to different kinds of energy injection processes. These lead to spectral distortions in the CMB such as the $\mu$ and $y$ distortions \cite{erler2018planck,chluba2012evolution,2014PTEP.2014fB107T,lucca2020synergy}. The CMB, with its anisotropies and distortions, provides a view into the history of our universe \cite{Cyr:2024sbd,chluba2016spectral,chluba2021new,shoemaker2016probing,Dodelson:2003ft,hu1995toward,hu1997cmb,hu1998complete}.

Axions and axion-like particles (ALPs) are hypothetical particles predicted by the Beyond Standard Model (BSM) theories \cite{dine1983not,preskill1983cosmology,Ghosh:2022rta,sakharov1994nonhomogeneity,sakharov1996large}. Peccei and Quinn proposed a solution to the strong-CP problem, which was identified to give rise to hypothetical particles named axions \cite{Peccei:1977ur,Wilczek:1977pj,Weinberg:1977ma,peccei1978short,peccei1986viable,Kim:1986ax,Cheng:1987gp,turner1990windows,berezhiani1991cosmology,peccei1996qcd,berezhiani2001strong,peccei2008strong,weinbergno}. Axions are produced as a result of the global Peccei-Quinn U(1) symmetry breaking by the vacuum expectation value of a scalar field \cite{peccei1986viable,Peccei:1977ur,peccei1996qcd}. Since axions and ALPs possess mass, they are Pseudo Nambu Goldstone Bosons (PNGBs) and make one of the dark matter candidates alongside Primordial Black Holes (PBHs), MAssive Compact Halo Objects (MACHOs), Weakly Interacting Massive Particles (WIMPs), sterile neutrinos, etc., and may only weakly interact with photons \cite{dine1983not,preskill1983cosmology,Ghosh:2022rta,sakharov1994nonhomogeneity,sakharov1996large,abbott1983cosmological,berezhiani1991cosmology,khlopov1999nonlinear,chadha2022axion}. 

Axions or ALPs show photon-axion oscillations, similar to the case of neutrino oscillations \cite{Raffelt:1996wa,Choi_2021,smirnov2005msw,duan2010collective,wolfenstein2018neutrino,bilenky1999phenomenology}. 
The strength of the photon-ALP interaction decides the probability of conversion of photons to ALPs in an ionized medium. This conversion can take place in the presence of magnetic fields, hence astrophysical systems are deemed to be sites for the production of ALPs from photons \cite{mcdonald2023axion,Raffelt:1996wa}. The existence of ALPs can thus be probed using galaxy clusters, neutron stars, as well as the Milky Way galactic halo \cite{lecce2025probing,conlon20143,carenza2021turbulent,simet2008milky,Mukherjee_2018,Mukherjee_2019,Mukherjee_2020,Mehta:2024pdz,Mehta:2024wfo,mondino2024axioninducedpatchyscreeningcosmic,day2016cosmic}.  The CMB photons as they travel through ionized plasma, acquire an effective mass, which if matches the ALP masses that exist in nature, will convert to ALPs with a probability depending on the strength of the photon-ALP interaction $g_{a\gamma}$, the transverse magnetic field, the electron density, and the photon frequency \cite{osti_22525054,Mukherjee_2018,Mukherjee_2020,Raffelt:1996wa}. This is called the resonant conversion, and the strength of such a conversion dominates over the non-resonant ones \cite{Mukherjee_2019}. 

Such a conversion can also take place in the Milky Way, as the photons travel through the ionized medium of the galactic halo in the presence of magnetic fields of the order of $\mu$G \cite{jansson2012galactic,jansson2012new,oppermann2012improved,adam2016planck}. 
This secondary anisotropy also leads to spectral distortions in the CMB. The spatial and spectral behaviour of the ALP signal is very different from that of CMB, and can be used to probe the signal \cite{Mukherjee_2020,Mehta:2024pdz,Mehta:2024sye,Mehta:2024wfo,mehta2025turbulence}. Thus, based on the magnetic fields and electron densities observed in the Milky Way, accompanied by the spectral behaviour of the ALP signal, ALPs of mass range $10^{-15}$ to $10^{-11}$ eV can be probed. For this analysis, we consider the phenomenon of CMB photon-ALP resonant conversion in the Milky Way \cite{Mukherjee_2018}. and show how the measurements can be improved over the current bounds from particle physics experiment CAST (CERN Axion Solar Telescope) at $g_{a\gamma} < 6.6 \times 10^{-11} \, \mathrm{GeV}^{-1}$ \cite{2017}.

Galactic signals can be probed using wide-field-of-view observations, as these mostly affect the CMB observations on large scales. 
LiteBIRD is an upcoming CMB space mission that will observe a large fraction of the sky with its wide field of view on fifteen frequency channels \cite{paoletti2024LiteBIRD,LiteBIRD2023probing,matsumura2014mission,takakura2023wide}. The presence of CMB to ALPs conversion in the presence of the galactic magnetic field can be probed by leveraging the nearly full-sky coverage and its multiple frequency coverage. 
The wide-field surveys enable the study of the galactic signals, such as the galactic synchrotron emission \cite{waelkens2009simulating,phillipps1981distribution}, dust emission \cite{finkbeiner1999extrapolation,benoit2004first}, etc., at large angular scales, as the Milky Way occupies a significant portion of the sky. The variation of the galactic ALP signal generated by the photon-ALP resonant conversion can also be studied at large angular scales, the strength and angular variation of which will depend highly on the Milky Way magnetic field and electron density profiles at those scales. 

The modeling of galactic profiles is difficult, due to various astrophysical and gravitational processes that take place, which may lead to turbulence in the electron densities and magnetic fields \cite{terry2001role,seta2021magnetic,haverkorn2014magnetic,beck2012origin,vallee2004cosmic,beck2013magnetic,unger2024coherent}. This uncertainty adds to the difficulty of finding the best-fit large-scale profiles from 3-dimensional (3-D) simulations,  which are computationally very costly. The galactic ALP signal shows non-Gaussian features \cite{mehta2025turbulence}, therefore, using the power spectrum to probe the signal is not enough and requires analyzing higher-order statistics.
We propose a much faster method to probe the galactic ALP signal using the spatial template matching technique. This technique uses the spatial variation of the galactic ALP signal in the CMB sky to probe the ALP coupling. The method involves finding the best-fit template (that can match the spatial features of the signal very well) from a large number of possible templates and using it to probe the ALP coupling. 

The paper is organized as follows. The ALP-photon conversion in the galactic halo is studied in Sec. \ref{sec:ALP_milkyway}. This is followed by the description of our template matching technique in Sec. \ref{sec:ALP_template}. The analysis using our technique is shown in Sec. \ref{sec:results}, followed by a summary of our work in Sec. \ref{sec:conclude}. We use natural units ($\hbar= 1$, $c= 1$, $k_B= 1$) in this work, if not otherwise stated.
In addition, this work uses the cosmological parameters from the results of Planck 2018 \cite{aghanim2020planck}.

\section{Photon to ALPs conversion in the Milky Way}
\label{sec:ALP_milkyway}
\subsection{Photon-ALP resonant conversion}

The weak photon-ALP interaction offers a means of probing ALPs through interconversions and ALP decay to photons. The various astrophysical objects, such as neutron stars, galaxies, galaxy clusters, etc., provide windows to ALPs of varying masses and coupling strengths \cite{lecce2025probing,conlon20143,carenza2021turbulent,simet2008milky,Mukherjee_2018,Mukherjee_2019,Mukherjee_2020,Mehta:2024pdz,Mehta:2024wfo,mondino2024axioninducedpatchyscreeningcosmic,day2016cosmic}. In the Milky Way, we can study ALPs in the mass range of approximately $10^{-11}$ eV to $10^{-15}$ eV.
The strength of the photon-ALP interaction is quantified by the photon-ALP coupling constant $g_{a\gamma}$. The photon-ALP interaction Lagrangian is given as
\cite{Raffelt:1996wa}:
\begin{equation}\label{eq:lagr}
  \mathcal{L}_{\mathrm{int}} = -\frac{g_{a \gamma} F_{\mu \nu}\tilde{F}^{\mu \nu} a}{4} = g_{a \gamma} \overrightarrow{E} \cdot \overrightarrow{B}_{\mathrm{ext}} a,
\end{equation}
where $F_{\mu \nu}$ being the electromagnetic field tensor, $\tilde{F}^{\mu \nu}$ its dual, and $a$ is the ALP field. The Lagrangian depends on the component of the electric field vector along the external magnetic field. For an unpolarized photon, the conversion results in a polarization of the photon perpendicular to the external magnetic field. The conversions take place when the resonant condition is met, i.e., the effective masses of photon and ALP are equal \cite{Mukherjee_2018}:
\begin{equation}
 m_a = m_{\gamma} = \frac{\hbar \omega_p}{c^2} \approx \frac{\hbar}{c^2}\sqrt{n_e e^2 / m_e \epsilon_{0}}, 
\label{eq:resonance mass}
\end{equation}
where $\omega_p$ is the plasma frequency, $n_e$ the resonant location electron density. The conversion strength depends on the photon-ALP oscillation length scale ($l_{\mathrm{osc}} = 2\pi / \Delta_{\mathrm{osc}}$), the electron density variation scale. These are quantified using the parameters:
\begin{equation}
\label{eq:lenparams}
\begin{split}
\Delta_a &= - m_a^2 / 2\omega  \,,
\qquad
\Delta_e \approx - \omega_p^2 / 2\omega \,,
\\
\Delta_{a\gamma} &= g_{a\gamma}B_{t} / 2 \,,
\qquad
\Delta_{\mathrm{osc}}^2 = (\Delta_a - \Delta_e)^2 + 4\Delta_{a\gamma}^2 ,
\end{split}
\end{equation}
where $B_t$ is the transverse magnetic field at the resonant location perpendicular to the propagation direction. 

The Milky Way electron density and magnetic field profiles are inhomogeneous and affect the photon-ALP conversion probability. For a linearly varying electron density, we use the Landau-Zener approximation, and the conversion probability is quantified using the adiabaticity parameter, which is given as \cite{Mukherjee_2020,Mehta:2024wfo,osti_22525054}:
\begin{equation}
P_{\mathrm{conv}} = 1 - e^{-\pi \gamma_{\mathrm{ad}}/2} \approx \pi \gamma_{\mathrm{ad}} / 2, 
\label{eq:prob}
\end{equation} 
where the adiabaticity parameter is:
\begin{equation}
\gamma_{\mathrm{ad}} = \frac{\Delta_{\mathrm{osc}}^2}{|\nabla \Delta_{e}|} = \left| \frac{2g_{a\gamma }^2 B_t^2 \omega (1 + z)}{\nabla \omega_p^2} \right|.
\label{eq:gamma_ad}
\end{equation}
The resulting photon will also be polarized perpendicular to the external magnetic field. 

For an inhomogeneous electron density profile, it is possible to have multiple resonances (say $N$ in number) \cite{mehta2025turbulence}. There will be a net conversion to ALPs for the odd number of conversions, the probability of which, to the leading order, along the line of sight, can be given as:
\begin{equation}
    P(\gamma \rightarrow a) \approx \frac{\pi}{2} \sum_{i = 1}^{N}  \gamma_{\mathrm{ad},i} \, ,
   \label{eq: Probab}
\end{equation}
The resulting polarization signal will depend on the superposition of the resulting photons from different resonant locations. Thus, there will be depolarization of the signal in the case of an incoherent magnetic field. 
In this analysis, the effects of Faraday rotation are neglected, as the phenomenon is weak at microwave wavelengths. That will otherwise cause a change in the polarization of the photons as they travel through the ionized plasma of the Milky Way \cite{Mukherjee_2018,Mukherjee_2020}.

\subsection{The spatial distribution of the electron density and magnetic field in the Milky Way}

The galactic ALP signal depends on the galactic electron density and magnetic field profiles. These are modeled using the cylindrical symmetry about the galactic plane, with the origin as the galactic center. Any point in the galactic system will be defined by the coordinates $r$, $\phi$ and $z$), where the radial distance from the galactic center parallel to the disk plane is $r$,
 and $z$ is the distance of the point from the plane of the galactic disk. The locations close to the galactic plane will be masked due to high foreground contamination.
 
\subsubsection*{Galactic Magnetic field:}
The large-scale galactic magnetic field depends on the contributions from the galactic disk and galactic halo \cite{jansson2012galactic,oppermann2012improved,adam2016planck}. We use the model of  Jansson et al. \cite{jansson2012new}, with scale height as revised by Gaensler et al. \cite{gaensler2008vertical}. The disk component transitions to the toroidal component of the magnetic field in the halo. 
This transition depends on the height $h_{\mathrm{disk}}$ and width $w_{\mathrm{disk}}$ of the transition. In addition, the toroidal field shows different magnitudes as well as different radial extents in the northern and southern parts of the galactic plane. The model used for the toroidal component of the magnetic field is given as:
\begin{equation}
B_{\phi}^{\mathrm{tor}} (r,z) = e^{-|z| / z_0} L(z,h_{\mathrm{disk}},w_{\mathrm{disk}}) \times 
\begin{cases}
    B_{n}(1 - L(r,r_n,w_h)), & \text{if $z>0$}.\\
       B_{s}(1 - L(r,r_s,w_h)), & \text{if $z<0$}.
  \end{cases}
    \label{eq:maggal}
\end{equation}
where $B_{n}$ and $B_{s}$ refer to the different magnetic field amplitudes, and $r_n$ and $r_s$ set the radial extent of the field in the northern and southern regions, respectively, while $w_h$ refers to the transition region width. The direction of the toroidal components in the two hemispheres is opposite and is incorporated using a negative sign for $B_s$. The values used for the parameters to generate the simulated signal are shown in Table \ref{tab:params}.

The large-scale cylindrical radial component of the magnetic field points radially outwards, while the axial component points away from the galactic plane. The poloidal component (vector sum of these components) is modeled. This component is axisymmetric, with the field direction given by the angle the magnetic field makes with the galactic mid-plane. This is given as the elevation angle $\Theta_X$, which is constant ($\Theta_X = \Theta_X^{0}$) for regions outside the galactocentric radius $r_{X}^c$. The amplitude of the field is given as:
\begin{equation}
B_{X}(r_p) = b_{X}e^{-r_p/r_X}\times 
\begin{cases}
    r_p/r, & \text{if $r > r_{X}^{c}$}.\\
       (r_p/r)^2, & \text{if $r < r_{X}^{c}$}.
  \end{cases},
    \label{eq:bx}
\end{equation}
where the field line passes through the mid-plane at a radius $r_p$ and $r_X$ is the scaling radius. The radius $r_p$ for the constant elevation region ($r > r_{X}^{c}$) is given as:
\begin{equation}
r_p = r - |z| / \tan{\Theta_X^{0}},
    \label{eq:rp_const}
\end{equation}
while for the varying elevation angle region ($r < r_{X}^{c}$),
\begin{equation}
r_p = \frac{rr_{X}^{c}}{r_{X}^c + |z|/\tan{\Theta_{X}^{0}}}.
    \label{eq:rp_const}
\end{equation}
The elevation angle variation in this region is given as:
\begin{equation}
\Theta_{X}(r,z) = \tan^{-1}\left( \frac{|z|}{r - r_p} \right).
    \label{eq:elevar}
\end{equation}
The radial magnetic field will then be given by $B_{r} = B_{X} \cos{\Theta_{X}}$, and the axial component as
$B_{z} = (|z| / z) B_{X} \sin{\Theta_{X}} $. The values used for the parameters to generate the simulated signal are shown in Table \ref{tab:params}.

{Although this model agrees well with the data, there remain several sources of uncertainty. The galactic magnetic field will, in general, have both regular and turbulent components with a varying contribution in the spiral arm of the galaxies, galactic halo, disks, and the local bubble. All of these contribute to the uncertainty in the inference of the magnetic field.  Along with these, the electron density modeling also depends on the relativistic and thermal electrons, which is also crucial in determining the magnetic fields \cite{jansson2012new,jansson2012galactic,oppermann2012improved,adam2016planck}. All these different uncertainties need to be mitigated in order to make a robust inference of the photon-ALP conversion signal.}

\begin{table}[htbp]
  \small
  \centering
  \caption{\textbf{Galactic profile parameters} }
  \subfloat[\textbf{Magnetic field  parameters}]{%
    \hspace{0.5cm}%
    \begin{tabular}{|c|c|}
        \hline
        Parameters & Fiducial values  \\
        \hline
        \hline
        $B_n$ & 1.4 $\mu$G \\
        \hline
        $B_s$ & -1.1  $\mu$G \\
        \hline
        $r_n$ & 9.22 kpc \\
        \hline
        $r_s$ & $> 16.7$ kpc  \\
        \hline
        $w_h$ & 0.2 kpc \\
        \hline
        $z_0$ & 5.3 kpc \\
        \hline
    $B_{X}$ & 4.6 $\mu$G \\
        \hline
         
$\Theta_{X}^{0}$ & $49 \, \mathrm{(in \, degrees)}$ \\
        \hline
$r_{X}^{c}$ & 4.8 kpc \\
        \hline
$r_{X}$ & 2.9 kpc \\
        \hline
    \end{tabular}%
    \hspace{.5cm}%
  } \hspace{1cm}
  \subfloat[\textbf{Electron density parameters}]{%
    \hspace{0.5cm}%
    \begin{tabular}{|c|c|}
     \hline
    Parameters & Fiducial values  \\
        \hline
        \hline
        $n_1$ & 0.035 $\mathrm{cm^{-3}}$ \\
        \hline
        $H_1$ & 1.8 kpc  \\
        \hline
        $A_1$ & 17 kpc\\
        \hline
        $n_2$ & 0.09 $\mathrm{cm^{-3}}$\\
        \hline
        $H_2$ & 0.14 kpc\\
        \hline
        $A_2$ &  3.7 kpc\\
        \hline
        $R_{\odot}$ &  8.5 kpc\\
        \hline
    \end{tabular}%
    \hspace{.5cm}%
  }\label{tab:params}
\end{table}

\subsubsection*{Galactic Electron density:}
 The galactic electron density gets a contribution from the thin and thick disks of the galaxy, the spiral arms, the interstellar medium (ISM), clumps, and voids. We use the electron densities corresponding to thin and thick disks as these are the significant contributors to the large-scale electron density \cite{cordes2004ne2001,miller2013structure,gaensler2008vertical}. {We use the Cordes et al. NE2001 model \cite{cordes2004ne2001}}, which includes the contribution of the local bubble to the electron density profile \cite{cox1997modeling,phillips1992electron,welsh2009trouble}. The net electron density is thus modeled as the sum of the contributions from the thin and thick disks  as follows:
\begin{equation}
 n_e^{thick}(r,z) = n_1\left[ \frac{\cos(\pi r/2A_1)}{\cos(\pi R_{\odot}/2A_1)}\right] \mathrm{sech}^2(z/H_1)U(A_1 - r), 
\end{equation}
\begin{equation}
 n_e^{thin}(r,z) = n_2 \exp\left[ -\frac{(r - A_2)^2}{A_2^2}\right] \mathrm{sech}^2(z/H_2)U(r) .
 \end{equation}
Here $\mathrm{sech}(x)$ refers to the hyperbolic secant function ($\mathrm{sech}(x) = 2 [e^{x} + e^{-x}]^{-1}$) which achieves the maximum value of 1 at $x = 0$ and tends to 0 at asymptotic values of $x \rightarrow \pm \infty$. The $\mathrm{sech^2}$ form is used as it facilitates a rounder variation of the electron density at $z=0$, as compared to an exponential dependence. The $U(r)$ denotes the step function ($=1$ for $r > 0$, and $=0$ for $r<0$), and the values of the parameters for the two components are shown in Table \ref{tab:params}. 

The electron density in the Milky Way will, in general, depend on the different components of the galaxy that contribute to the electron density. This includes the spirals, galactic halo, disks, the local bubble, voids, clumps, etc., that need to be accounted for. The Faraday Rotation Measures (RMs) and Dispersion Measures (DMs) of pulsar observations are used to model the galactic electron density. The precise modeling would require observing a large number of pulsars and adopting a realistic location for the solar system, with respect to the galactic plane. The uneven distribution of a finite number of pulsars in the galactic halo, based on their luminosities, leads to some regions being not so well modeled as others. The turbulence in various components will affect the electron density at small scales. In addition, the presence of HII regions and supernova remnants adds further uncertainty to the modeling of electron density. The modeling assumes some form of symmetry, which is not the case for turbulent profiles \cite{cordes2004ne2001,miller2013structure,gaensler2008vertical,cox1997modeling,phillips1992electron,welsh2009trouble}. So, for searches of ALPs signal from Milky Way, it is crucial to understand the impact of mis-modeling of electron density and how it can impact the inference of the signal.

\subsection{Simulating the ALP signal in the Milky Way}
\label{sec:simalp}
\begin{figure}[h!] 
\centering
\includegraphics[height=7cm,width=11cm]{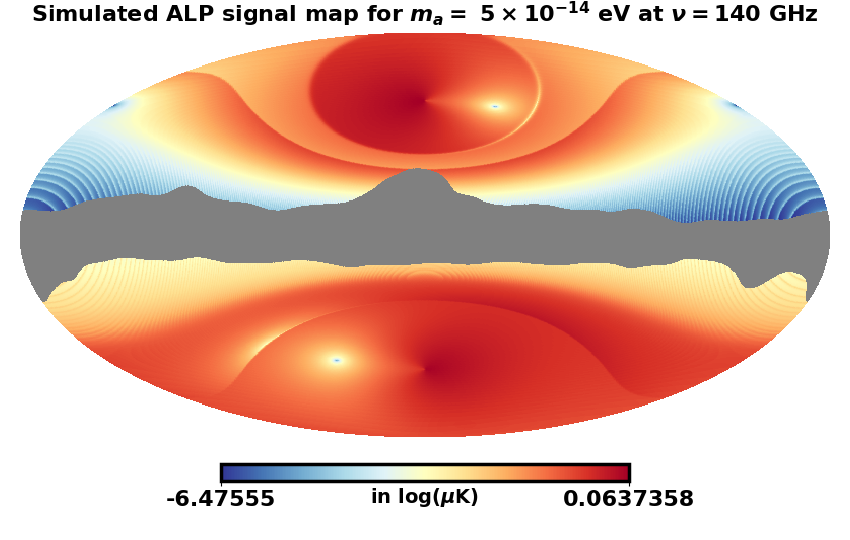}
\caption{Simulated galactic ALP polarized intensity signal for $m_a = 5\times 10^{-14}$ eV at 140 GHz ($g_{a\gamma} = 3 \times 10^{-11} \, \mathrm{GeV}^{-1}$).} 
\label{fig:sim4} 
\end{figure}

\begin{figure}[h!] 
\centering
\includegraphics[height=7cm,width=11cm]{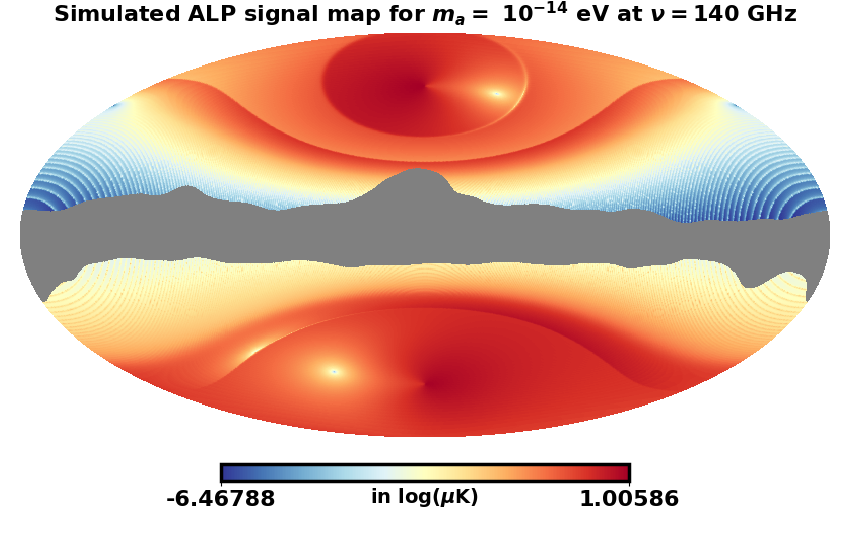}
\caption{Simulated galactic polarized intensity ALP signal for $m_a = 10^{-14}$ eV at 140 GHz ($g_{a\gamma} = 3 \times 10^{-11} \, \mathrm{GeV}^{-1}$).} 
\label{fig:sim5} 
\end{figure}

\begin{figure}[h!] 
\centering
\includegraphics[height=7cm,width=11cm]{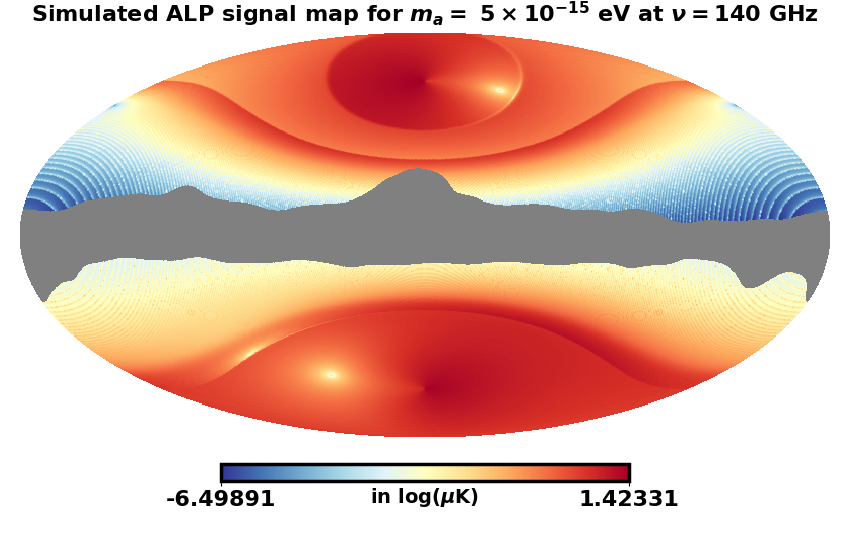}
\caption{Simulated galactic ALP polarized intensity signal for $m_a = 5\times 10^{-15}$ eV at 140 GHz ($g_{a\gamma} = 3 \times 10^{-11} \, \mathrm{GeV}^{-1}$).} 
\label{fig:sim6} 
\end{figure}

The galactic ALP signal will affect the CMB in both the temperature and polarization \cite{Austermann_2012,2020_planck}. The ALP signal along the line of sight in temperature depends on the transverse magnetic field strength $B_t$ and causes a loss in CMB intensity, as there is a loss of photons if they convert to ALPs. The transverse magnetic field strength is given as:
\begin{equation}
B_t = \sqrt{B_{\theta}^2 + B_{\phi}^2 }.
    \label{eq:Bt}
\end{equation}
The polarization signal is determined by the magnitude and direction of the transverse magnetic field at the resonant location. Since the lines of sight point to spherically radial directions ($\hat{\rho}$), the polarization is determined by the relative magnitudes of the components $B_{\theta}$ and $B_{\phi}$. The total intensity of the ALP signal along a line of sight is calculated approximately as the sum of the distortion intensities of individual resonances (say, $N$ in number):

\begin{equation}
    \Delta I_T^{\alpha}(\nu) = \frac{\pi}{2} \sum_{i = 1}^{N}  \gamma_{\mathrm{ad},i} \, I_0 (\nu) \, ,
   \label{eq: tempalp}
\end{equation}
where $I_0 (\nu)$ is the CMB black-body intensity. 
The net polarization signal is obtained by the superposition of photon polarization signals from various resonances along a line of sight. The polarization can be written in terms of the Stokes parameters Q and U \cite{ghatak2009optics,Austermann_2012,2020_planck}. The corresponding signals are given as:
\begin{equation}
    \Delta I_Q^{\alpha}(\nu) =  \frac{\pi}{2} \sum_{i = 1}^{N}  \gamma_{\mathrm{ad},i} (\cos{2\beta})_i \, I_0 (\nu) \, ,
   \label{eq: tempalp}
\end{equation}

\begin{equation}
    \Delta I_U^{\alpha}(\nu) =  \frac{\pi}{2} \sum_{i = 1}^{N}  \gamma_{\mathrm{ad},i} (\sin{2\beta})_i \, I_0 (\nu) \, ,
   \label{eq: tempalp}
\end{equation}
 where $\beta$ is the angle the transverse magnetic field makes with respect to the $\hat{\phi}$ direction. The polarized intensity map is obtained as:
\begin{equation}
    \Delta I_{P}^{\alpha} (\nu) =  \sqrt{\Delta I_Q^{\alpha}(\nu)^2 + \Delta I_U^{\alpha}(\nu)^2} \, .
   \label{eq: tempalp}
\end{equation}
The intensity values are converted to the CMB temperature units using the $dI/dT$ derivative for the CMB black-body:
\begin{equation}
    \Delta T^{\alpha} = \left(\frac{dI}{dT}\right)_{\mathrm{cmb}}^{-1} \Delta I^{\alpha} \, .
\end{equation}

The difference in powers between the temperature and polarization signals depends on the number of resonant locations and the magnetic field coherence lengths, as depolarization is high for a large number of resonances and a low magnetic field coherence scale (resulting in random orientation of the magnetic field directions). The magnetic field in the galaxy can be quite turbulent, leading to low coherence scales, but the number of resonances will restrict this depolarization \cite{mehta2025turbulence}. {There can be multiple resonances that can lead to a high depolarization of the polarization signal, but will be accompanied by an increase in the ALP signal with temperature, due to a loss in the CMB intensity at each resonance.}

Using the formalism describe above, we can model the photon to ALPs conversion signal for different ALP masses and coupling strengths. We show a few representative simulations of the polarization maps for ALP masses $5 \times 10^{-14}$ eV, $10^{-14}$ eV and $5 \times 10^{-15}$ eV using the mentioned magnetic field and electron density models and coordinate transforming them to spherical coordinates, with the solar system as the origin. The observation point is the solar system, which is at the distance of about 8.5 kpc from the galactic center. To obtain the signal from different directions at this observation point, we coordinate transform to the solar system frame which is the origin. This is followed by an additional coordinate transform from cylindrical to spherical coordinates.
This is needed so that the magnetic field and electron densities along each line of sight (corresponding to the spherical radial  direction) can be obtained. We use the HealPIX NSIDE = 512  and the LiteBIRD frequency of 140 GHz to simulate the maps. We mask 20\% of the sky map, corresponding to latitudes around the galactic plane. We use a domain size of 150 pc for the profiles \cite{terry2001role,seta2021magnetic,haverkorn2014magnetic,beck2012origin,vallee2004cosmic,beck2013magnetic,unger2024coherent}.  The maps are smoothed to the beam resolution corresponding to this frequency channel and are shown in Fig. \ref{fig:sim4} for $m_a = 5\times 10^{-14}$ eV, in Fig. \ref{fig:sim5} for $m_a =  10^{-14}$ eV, and in Fig. \ref{fig:sim6} for $m_a = 5\times 10^{-15}$ eV.  
Generally, the ALP signal increases with decreasing ALP mass \cite{Mukherjee_2018}. Also, the features of the map change with the mass of ALPs, due to changing magnetic field and electron density at the resonant locations.  The high mass ALPs form nearer to the galactic center than do low mass ALPs, and that can be seen from the position of the ring-like features above and below the midplane. The north-south asymmetry is mainly a result of the magnetic field strength asymmetry, with higher signal from the southern hemisphere owing to stronger magnetic fields in the southern hemisphere \cite{jansson2012galactic,jansson2012new,oppermann2012improved,adam2016planck}. Additionally, there are local features near the ring-like features: cold spots are to the right part of the rings in the northern hemisphere, and to the left half in the southern hemisphere. This can be explained as a result of the toroidal magnetic field about the galactic center, which, after coordinate transforming to the solar system coordinates, results in a decrease in the transverse magnetic field along the line of sight. 

{Spectrally, the ALP signal will show a distinct frequency variation, and the multi-frequency maps will be spatially correlated with the variation in magnetic field and electron density at different scales and locations. The fluctuations due to the ALP distortion signal in different locations in the sky will be highly non-Gaussian \cite{mehta2025turbulence}, capturing the variation in the galactic profiles. The large-scale coherent field will lead to the large-scale variation in the spatial shape of the ALP signal, while turbulence will lead to small-scale features, which are difficult to model. The galactic profiles will, in general, show a combination of both coherent and turbulent components. This variation can only be captured by higher-order statistics, calling for analysis beyond the power spectrum. In the next section, we introduce a sky-template-based  technique that avoids this problem by trying to capture the spatial large-scale behavior of the signal at the map level.}

\section{A Novel Method to infer the 
Galactic ALPs signal from CMB Observations}
\label{sec:ALP_template}
\subsection{Why a template-based approach to search for ALP signal from the Milky Way?}
Traditionally, various CMB anisotropies have been probed using their spatial variation at different angular scales. This calls for a search based on the power spectrum, which is a harmonic transform of the two-point correlation function of the anisotropic CMB at different locations \cite{Dodelson:2003ft}. This method is efficient when the signal obeys Gaussian statistics.
The galactic ALP distortion signal is highly non-Gaussian, the sky-projected profile of which follows the variation in electron density and magnetic field profiles \cite{mehta2025turbulence}. A power-spectrum-based approach will not be able to capture the non-Gaussian aspect of the signal. Moreover, there are galactic dust and synchrotron contamination, which will affect the estimation of the signal from power spectrum or higher order spectra \cite{aghanim1999searching,hanson2009estimators,challinor2012cmb,liguori2006testing,ade2016planck}. 

The photon-ALP resonant conversion is a local phenomenon and will depend on the electron density and magnetic field variation at small scales. The galactic profiles may  be quite turbulent due to a number of astrophysical processes in the interstellar medium, and any measurement on them from multi-band observations will be difficult. This leads to modeling uncertainties around these profiles and a wrong estimation of the ALP signal \cite{mehta2025turbulence}. To go around this problem would require us to find the best-fit electron density and magnetic field profiles at large scales from an ensemble of possible sets of electron density and magnetic field profiles, which can capture the large-scale profile. The ALPs signal at the large angular scales will depend primarily on the large-scale electron density and coherent magnetic field. The small-scale magnetic field and electron density will be more uncertain, and modeling those will be more difficult. This segregation of the large and small scales can help in building a spatial template-based search technique that can capture only the large-angular-scale effect and mitigate the small-scale features. Although the small-scale features can be very informative in mapping the ALP signal, it is more prone to modeling systematics. As a result, we develop a method that can use the signal in those scales that are more reliable to infer the ALP signal. Moreover, the estimation of the
ALP signal requires the calculation of conversion probabilities at various resonant locations in the 3-D galactic halo for a large portion of the sky. Such a search that involves 3-D calculations for the best-fit profile from millions of test profiles  
will be computationally very costly. This calls for an approach that is computationally inexpensive and efficient amid the uncertainties surrounding the galactic profiles.

We propose a new way of probing the photon-ALP conversion in the Milky Way using a spatial template matching technique at every frequency channel accessible from a CMB experiment. This method captures the sky projected 2-dimensional (2-D) structure of the signal on large scales, and does not require solving for the 3-D electron density and magnetic field profile information to search for an ALP signal from the data. As a result, this will enable a computationally inexpensive estimation of the ALP coupling from the best-fit sky template that captures the large-scale variation of the galactic ALP signal in terms of a few effective parameters, which capture the large-scale features of the signal. It is important to point out that the modeling of the signal in 2-D for only the large scales loses a lot of map-level information content at small scales. However, the small-scale features are more difficult to model due to inaccuracy in the electron density and magnetic field uncertainty. As a result, the proposed 2-D template-based technique is useful for a fast inference of the signal from the map. But if one finds any non-zero evidence of a signal in favor for ALPs, then a more detailed modeling of 3-D magnetic field and electron density will be necessary to also fit the small-scale features of the signal and enhance the signal-to-noise ratio of the detection. 

{The fluctuations of the ALP signal at different locations in the sky (represented by different pixels of the observing survey) when stacked together will show a non-Gaussian distribution, compared to the Gaussian distribution for an only CMB map. The non-Gaussianity of the ALP signal will depend on the underlying galactic electron density and magnetic field profiles, and the related small-scale turbulence in them. The sky template-based technique does not use the small-scale information about these profiles, and ensures that the large-scale non-Gaussian nature of the signal is used in different frequency channels to infer the ALP coupling strength using almost the full sky signal in temperature and polarization.} {The spatial template will be the same in all frequency channels, and the template maps will follow the spectral dependence of the ALP signal. This spectral and spatial information ensures the robustness of the method, which can be applied to other galactic signals across the sky as well, such as the diffused synchrotron emission \cite{waelkens2009simulating,phillipps1981distribution}, dust emission \cite{finkbeiner1999extrapolation,benoit2004first}, etc.}


\subsection{Sky template for the ALP signal}
{The projected ALP signal in polarization may differ depending on the number of resonant locations along various lines of sight and the accompanying depolarization \cite{mehta2025turbulence}, depending on the shape and turbulence in the electron density and magnetic field profiles. A large number of resonances may lead to high depolarization and a large-scale variation in the shape of the ALP signal, but that will lead to an increase in the ALP temperature signal, which will add up, corresponding to a loss in intensity of the CMB. So, we can probe both the temperature and the polarization distortion of the CMB to discover the signature of ALPs.}  

Modeling the ALP signal sky requires us to find an element of symmetry, which, for the galactic plane, is the perpendicular axis through the galactic center. We will thus model the signal first around this axis through the galactic center and then coordinate transform the signal to solar system coordinates (at a distance of 8.5 kpc from the galactic center) to get a model map of the signal with the center as the solar system. For a large-scale electron density variation, we 
model the polarized intensity map. We do not model the Q and U maps separately, but form templates of the polarized intensity maps to keep our template approach independent of the local orientation of the galactic magnetic field lines.

\subsubsection{Latitude-based modelling of the ALP signal at the Galactic center}
In this section, we represent quantities in the solar system frame with an overhead asterisk ($^*$), while those in the galactic center frame without it. Assuming a cylindrical symmetry about the perpendicular axis through the galactic center, the large-scale electron density and magnetic field at a point in the galactic halo are determined by the distance of the point from the plane of the galactic disk ($z$) and the radial distance from the galactic center parallel to the disk plane.  
The net effective ALP signal along a line of sight ($\theta^{*},\phi^{*}$) in the solar system coordinates will be a result of the signal being generated at multiple resonances, depending on the turbulence in galactic profiles. 
With the origin as the galactic center, we treat the net polarization signal ($\Delta I$) being generated at an effective radial distance from the perpendicular axis through the galactic center ($r$), which for a cylindrically symmetric signal should be the same at all points at a certain height ($z$) from the galactic plane. All these points will correspond to a certain latitude ($\theta$) in spherical coordinates (as $\tan{\theta} = z / r$). We thus try to model the relation between $r$ and $\theta$ using different class of symmetric functions $f$ about the latitude corresponding to the galactic plane at $\theta = \pi/2$:
\begin{equation}
r = f(d,\theta), 
    \label{eq:temp_r}
\end{equation}
where $d$ is a length scale corresponding to the function that sets the maximum value of the effective radial distance $r$. 
The electron density decides where the ALP signal is generated based on the resonance condition being met ($m_a = m_{\gamma} \propto n_e$). On large scales, the electron density is expected to be higher near the galactic plane (at $\theta = \pi/2$) and closer to the central axis (for low values of $r$). 
We thus select functions $f$ such that $r$ peaks at the galactic plane ($\theta = \pi/2$) and decays near the poles. 

\begin{figure}[h!] 
\centering
\includegraphics[height=7cm,width=12cm]{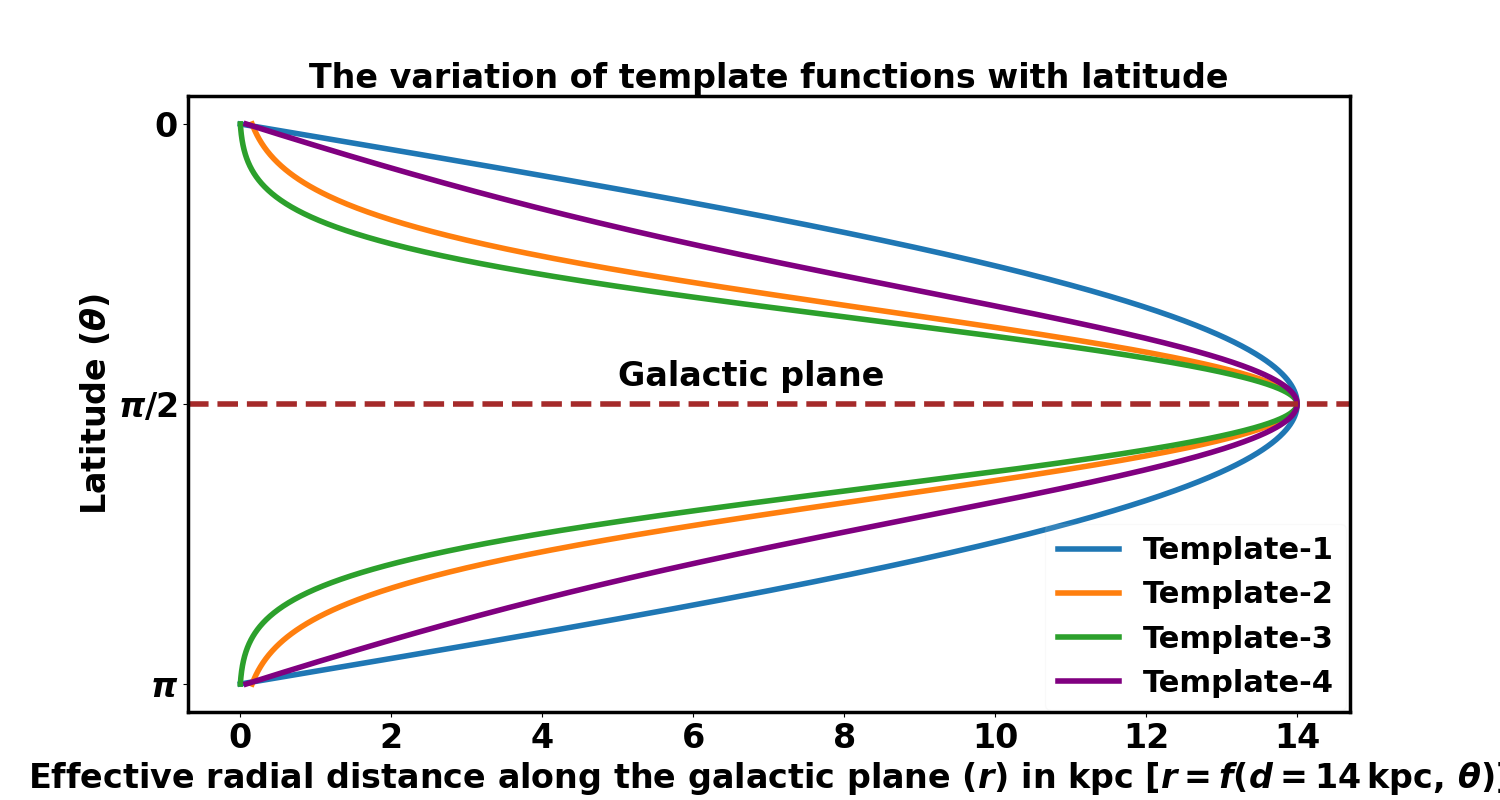}
\caption{Variation of template functions with latitude. The galactic plane is along the latitude $\theta = \pi/2$. The function decays near the poles. Here $d = 14$ kpc.} 
\label{fig:funcs} 
\end{figure}

\begin{table}[h!]
\centering

\begin{tabular}{|c|c|}

\hline

\textit{\textbf{Template Name}} &
\textit{\textbf{Functions ($f(d,\theta)$)}}
\tabularnewline \hline
Template-1 & $d \sin{\theta}$ \\
 \hline
Template-2 & $d\exp[{- (\theta - \pi/2)^2 / 2\sigma^2], \, \, \sigma= \pi/6}$  \\
 \hline
Template-3&$d \sin{\theta}\exp[- (\theta - \pi/2)^2 / 2\sigma^2], \, \, \sigma= \pi/6 $ \\
 \hline
Template-4& $0.5 \, d \sin{\theta} + 0.5 \, d \exp[- (\theta - \pi/2)^2 / 2\sigma^2] , \, \, \sigma= \pi/6 $ \\
 \hline
 
\end{tabular}
\caption{Model functions and their labels.}
\label{tab:dfunc}
\end{table}
We use the template functions as shown in Fig. \ref{fig:funcs} with the galactic plane along the latitude $\theta = \pi/2$. The template functions used are provided in Table \ref{tab:dfunc} and point to different variations of $r = f(d,\theta)$ with latitude, all of which decay near the poles. These effective radial distance values $r(\theta)$ for a certain $d$ corresponding to different latitudes $\theta$ will enable us to perform a coordinate transform to the solar system frame. In the next section, we will coordinate transform to the solar system frame to obtain template maps corresponding to different template functions ($f$), for a certain maximum effective length scale ($d$).
\subsubsection{Modeling the signal at the solar system}

A latitude-based coordinate transform to the solar system frame would map various sky locations ($\theta$, $\phi$) in the galactic center frame to points in the frame of the solar system ($\theta^{*}$, $\phi^*$). Thus, we model the ALP signal simulated intensity in the solar system frame, i.e. $\Delta I(\theta^{*}$, $\phi^{*})$ in bins of latitudes of width $\Delta \theta$ around $\theta$ in the galactic center frame by using the effective radial distances $r(\theta)$ from Eq. \eqref{eq:temp_r}, and coordinate transform to the solar system frame.

The net amplitude of the signal at a latitude in the galactic center frame can be expected to be a constant term $A$, using the large-scale symmetry of the profiles about this point. 
But this would not capture the left-right asymmetry (as explained in Sec. \ref{sec:simalp}), generated due to the differing contributions of the azimuthally symmetric toroidal component (at the galactic center) to the transverse magnetic field along the line of sight, when coordinate-transformed to the solar system as the origin. Thus, we include an additional term that depends on the direction being observed. The net signal at a sky location can then be modelled as:
\begin{equation}
\Delta I (\theta^{*}, \phi^{*}) = A (\theta) + B (\theta)  \cos{\theta} \sin{\phi}.
    \label{eq:temp_amp}
\end{equation}
We fit the ALP simulated signal (see Sec. \ref{sec:ALP_milkyway}), which is obtained by calculating the photon-ALP conversion probability along each line of sight, based on the 3-D electron density and magnetic field profiles. The signal will vary for different mass ALPs and we model the ALP simulated signal maps with the template functions given in Table \ref{tab:dfunc}, with the intensity in the solar system frame  given by Eq. \eqref{eq:temp_amp}.  Thus, we model the intensity at the observing sky location ($\theta ^*$, $\phi ^*$) using SciPy's curve\_fit function \cite{2020SciPy-NMeth}. This is done by coordinate transforming for each latitude bin $\Delta \theta$ around the galactic center frame latitude $\theta$ to the solar system frame using the respective effective radial distance values $r(\theta) = f(d,\theta)$ for a given $d$. This gives us a template map corresponding to the ALP signal simulated map. The various best-fit template maps obtained from different functions are analyzed in Appendix \ref{sec:compare}. The quality of the fit will depend on how well the intensity coefficients $A(\theta)$ and $B(\theta)$ model the ALP simulated intensity at those latitudes. The effective radial distance ($r(\theta)$) also determines the quality of fit by mapping the variation of the resonant locations with latitude. 

The net intensity we model in the solar system frame at the sky-location ($\theta^*$,  $\phi^{*}$) is mapped to ($\theta$, $\phi$) in the galactic center coordinates. We set the condition that $A > 0$ and $|B| < |A|$ to capture the small-scale features using the second term with coefficient $B$, and ensure that the intensity values stay positive. Using a latitude-based coordinate transform would map the intensity values in the galactic center frame to the solar system frame using the effective radial distance $r$, calculated from Eq. \eqref{eq:temp_r}. 
 
We form non-overlapping slices of latitudes of small widths $\Delta \theta$ around $\theta$ in the galactic coordinate frame.  The effective radial distance $r$ of the signal in the galactic center frame is calculated for a given  $f$ using Eq. \eqref{eq:temp_r}. Shifting to the solar system frame will locate the signal in the solar system coordinates.  
 The net modeled signal is calculated using Eq. \eqref{eq:temp_amp} and mapped to points in the solar system frame.
The points that lie within the same latitude slice with mean $\theta$ in the galactic center coordinates are modeled with the same parameters $A(\theta)$ and $B(\theta)$, where $\theta$ denotes the latitude corresponding to the solar system coordinates. A template is thus obtained, which for a pixelized map may be missing some values, which can be obtained by interpolation of the signal pixels on the map. This method is followed to generate the polarized intensity template map.





\section{Application of the technique using LiteBIRD instrument specification}
\label{sec:results}

\begin{table}[h!]
\centering

\begin{tabular}{|c|c|c|}

\hline

\textit{\textbf{Frequency}} &
\textit{\textbf{Noise (in $\mu$K-arcmin)}}&
\textit{\textbf{Beam (FWHM in arcmin)}}
\tabularnewline \hline
60 & 21.31 &51.1 \\
 \hline
68 & 19.91 & 41.6 \\
 \hline
78 & 15.55 & 36.9
\\
 \hline
89 & 12.28 & 33.0 
 \\ 
 \hline
100 & 10.34 & 30.2 
 \\ 
 \hline
119 & 7.69 & 26.3
 \\ 
 \hline
140 & 7.25 & 23.7 
 \\ 
 \hline
166 & 5.57 & 28.9 
 \\ 
 \hline
195 & 7.05 & 28.0 
 \\ 
 \hline
235 & 10.79 & 24.7 
 \\ 
 \hline
280 & 13.80 &  
 22.5 \\ 
 \hline

\end{tabular}
\caption{LiteBIRD frequency channels used with their beams and noises.}
\label{tab:liteb}
\end{table}
We model the simulated maps for various ALP masses ($5 \times 10^{-14}$ eV, $ 10^{-14}$ eV and $5 \times 10^{-15}$ eV) using different functions $f(d,\theta)$, with various values of the length scale $d$. For each case, we use the latitude width of $\Delta \theta = 0.05 $ rad in the galactic center coordinates. We vary $d$ from 12 to 20.5 kpc in steps of 0.5 kpc. The functions $f(d,\theta)$ we use are shown in Table \ref{tab:dfunc}. The functions are chosen such that they tend to 0 at the poles, corresponding to $\theta = 0 $ and $\theta = \pi$. These give us a number of template maps for each simulated map. We compare the template maps using various functions in Appendix \ref{sec:compare}. Since there can be a number of resonances along a line of sight, the models capture the variation in the effect resonant location at a certain latitude $\theta$. The maximum distance of this variation is taken along the galactic plane and (at $\theta = \pi/2$) corresponds to the value $d$. Template-1 provides a slow variation of the effective resonant location with latitude away from the galactic plane, followed by a steep one, while Template-3 provides a steep variation, followed by a slow one. Template-2 and Template-4 lie between the two extremes, with Template-4 being the average of Template-1 and Template-2.   We model our maps at the LiteBIRD frequency of 140 GHz and set a flux threshold of $10^{-3} \, \mu$K, which implies that we do not consider pixels below this flux threshold value in our analysis. We model the signal at different frequencies using the same spatial template, but with different spectral behavior as described in Sec. \ref{sec:ALP_milkyway}. 

 In an observed frequency map, the ALP signal ($A$) will be contaminated by CMB ($C$), which is independent of frequency, while foregrounds ($S_f$) and instrument noise ($N$) will be frequency dependent. The mock data map can then be written as:
\begin{equation}
S(\nu) = C + A(\nu) + S_f(\nu) + N(\nu).
    \label{eq:obs_map}
\end{equation}
 The simulated maps are obtained by combining the CMB, simulated ALP signal, galactic synchrotron (model "s3" from PySM \cite{Thorne_2017}), thermal dust (model "d3" from PySM), smoothed by the beams corresponding to the respective frequency channels. The instrument noise corresponding to the frequency channels is added. This gives us the mock data maps at different frequency channels accessible from LiteBIRD.

\subsection*{Cleaning of the ALP signal:}
The galactic ALP polarization signal will be contaminated by galactic foregrounds  (thermal dust and synchrotron), as well as lensed CMB anisotropy \cite{dickinson2003towards,dickinson2016cmb,ichiki2014cmb,carretti2010galactic,tegmark2003high}. These signals can be separated using spectral cleaning techniques by using the large number of channels available with LiteBIRD which covers the frequency range from 40 GHz to 450 GHz. Thus, LiteBIRD will provide an opportunity to probe the ALP signal with its wide field of view and numerous frequency channels \cite{paoletti2024LiteBIRD,LiteBIRD2023probing,matsumura2014mission,takakura2023wide}. 
We use eleven LiteBIRD frequency channels from 60 GHz to 280 GHz. The corresponding beams and noises are shown in Table \ref{tab:liteb}.

The contamination from foregrounds can be reduced by masking the galactic plane. We mask 20\% of the sky map, corresponding to the latitudes around the galactic plane. But that is not sufficient, as the contamination still remains away from the galactic plane.
Thus, a cleaning of the signal is required to diminish the effects of these components on the ALP signal. The maps are smoothed to the lowest resolution of 51.1 arcminutes, corresponding to LiteBIRD's 60 GHz frequency channel. We use the Interior Linear Combination (ILC) method, which uses a weighted linear combination of multi-frequency maps at a common resolution to make a cleaned map \cite{ilc2008internal,Eriksen_2004}. The large number of frequency channels that LiteBIRD will use to observe the sky will be very efficient in cleaning the ALP signal from various contaminants \cite{paoletti2024LiteBIRD,LiteBIRD2023probing,matsumura2014mission,takakura2023wide,Hansen_2006}. The weights are obtained using the covariance of the observed  polarization maps ($C_S$) and are given as:
\begin{equation}
w_{\rm{ilc}} = f_{\alpha}^T C_S^{-1} (f_{\alpha}^T C_S^{-1} f_{\alpha})^{-1},
    \label{eq:cov}
\end{equation}
where the spectral dependence $f_{\alpha}$ of the ALP signal at the intensity level goes as $f_{\alpha} \propto \nu I_0(\nu)$, where $I_0(\nu)$ is the CMB black-body intensity spectrum. Using these weights, the ILC-cleaned map is obtained as:
\begin{equation}
S_{\rm{ilc}} = \sum_{\nu} w^{\nu} S(\nu). 
    \label{eq:ilcmap}
\end{equation}

  The same weights are used to find the cleaned template ($M_{\rm{ilc}} = \sum_{\nu} w^{\nu} M(\nu)$) using the multi-frequency template maps, smoothed to the same common resolution. To get the covariance corresponding to the ILC-cleaned map, 500 realizations of the frequency maps are made and ILC-cleaned.

 In the following sections, we will use the term mock data map to mean the ILC-cleaned mock data map ($S_{\rm{ilc}}$). The term template would mean the multi-frequency templates combined with the respective ILC weights ($M_{\rm{ilc}}$). Similarly, the covariance of the maps would mean the covariance corresponding to the ILC-weighted combination of multi-frequency maps without the ALP signal.
\subsection{Finding the best-fit model}
\label{sec:bestfit}



\begin{figure}[h!]
     \centering
     \begin{subfigure}[h]{0.45\textwidth}
         \centering    
\includegraphics[height=5cm,width=7.3cm]
         {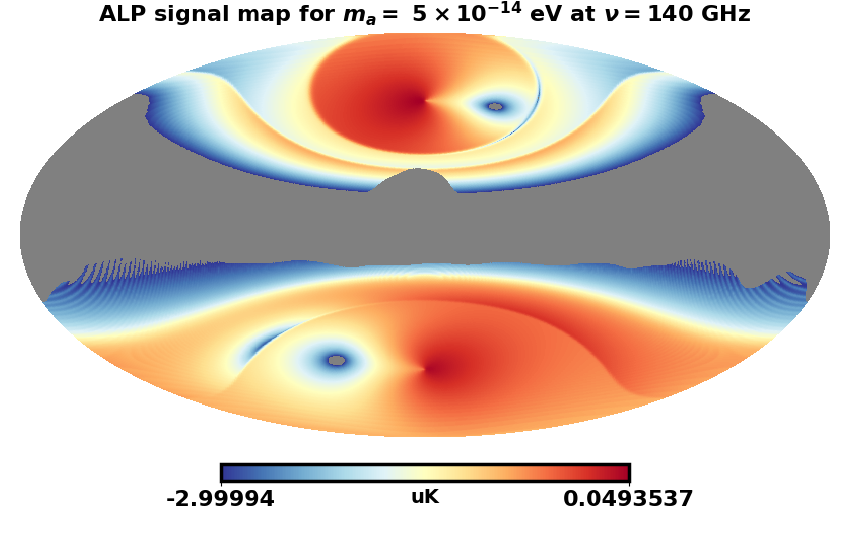}
         \caption{{Galactic ALP signal map.}}
         \label{fig:sig_4}
     \end{subfigure} 
     \hspace{0.01cm} 
     \begin{subfigure}[h]{0.45\textwidth}
         \centering
    \includegraphics[height=5cm,width=7.3cm]
         {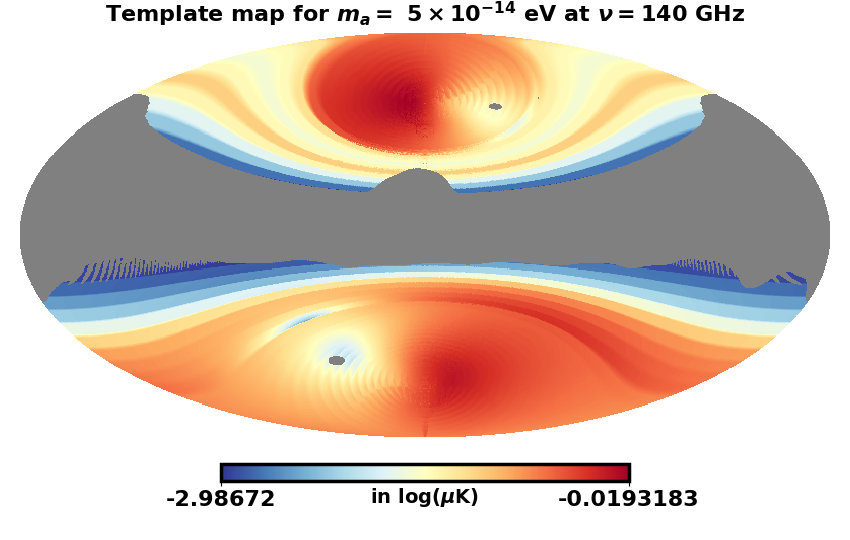}
         \caption{{Best-fit template map.}}
         \label{fig:mod_4}
     \end{subfigure}
\caption{ALP signal (smoothed to 51.1 arcmin) and best-fit template maps for $m_a = 5\times 10^{-14}$ eV at $\nu =$ 140 GHz ($g_{a\gamma} = 10^{-11} \, \mathrm{GeV^{-1}}$). }

\label{fig: sigmod_4}
\end{figure}

\begin{figure}[h!]
     \centering
     \begin{subfigure}[h]{0.45\textwidth}
         \centering    
\includegraphics[height=5cm,width=7.3cm]
         {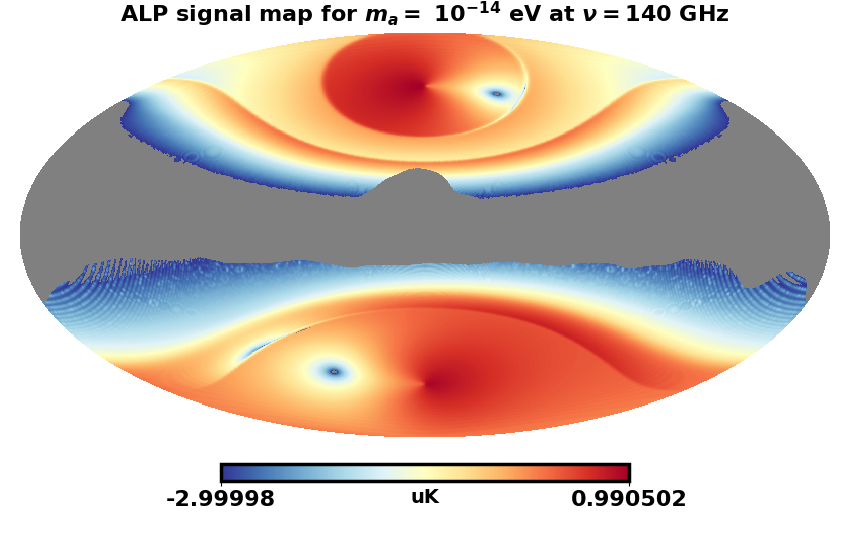}
         \caption{{Galactic ALP signal map.}}
         \label{fig:sig_5}
     \end{subfigure} 
     \hspace{0.01cm} 
     \begin{subfigure}[h]{0.45\textwidth}
         \centering
    \includegraphics[height=5cm,width=7.3cm]
         {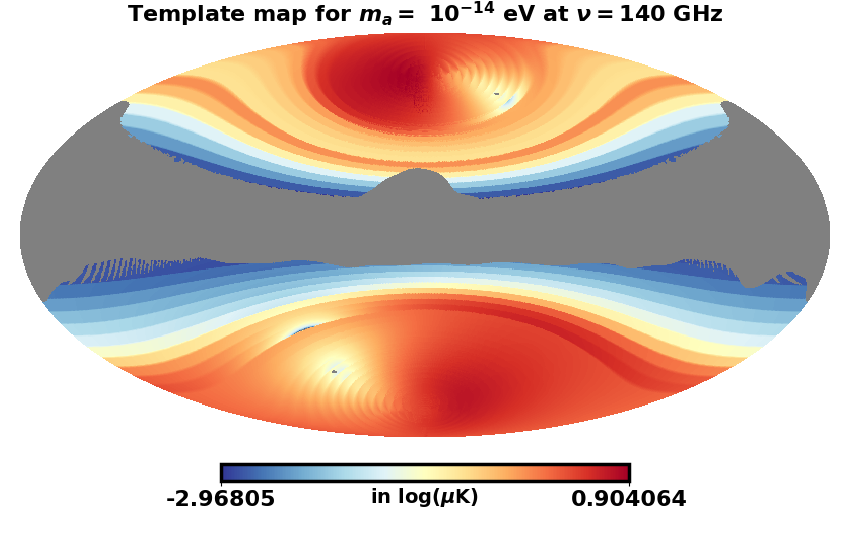}
         \caption{{Best-fit template map.}}
         \label{fig:mod_5}
     \end{subfigure}
\caption{ALP signal (smoothed to 51.1 arcmin) and best-fit template maps for $m_a =  10^{-14}$ eV at $\nu =$ 140 GHz ($g_{a\gamma} = 10^{-11} \, \mathrm{GeV^{-1}}$). }

\label{fig: sigmod_5}
\end{figure}

\begin{figure}[h!]
     \centering
     \begin{subfigure}[h]{0.45\textwidth}
         \centering    
\includegraphics[height=5cm,width=7.3cm]
         {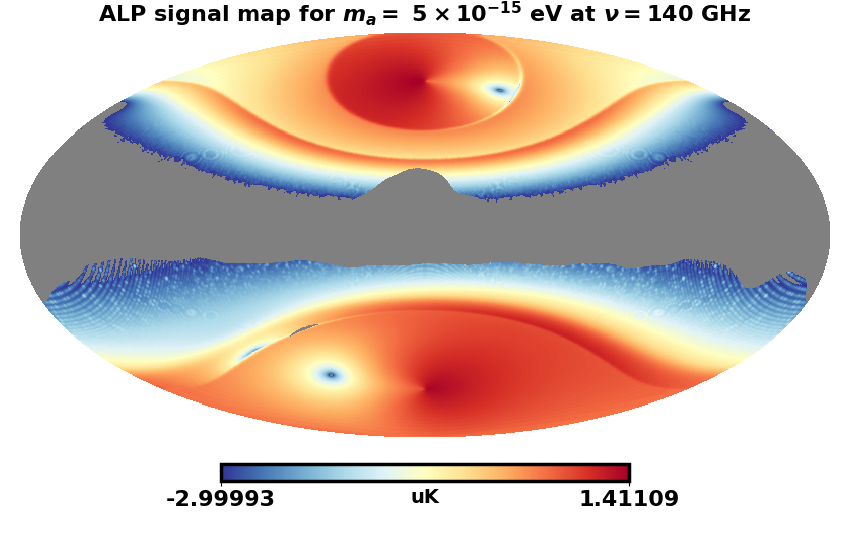}
         \caption{{Galactic ALP signal map.}}
         \label{fig:sig_6}
     \end{subfigure} 
     \hspace{0.01cm} 
     \begin{subfigure}[h]{0.45\textwidth}
         \centering
    \includegraphics[height=5cm,width=7.3cm]
         {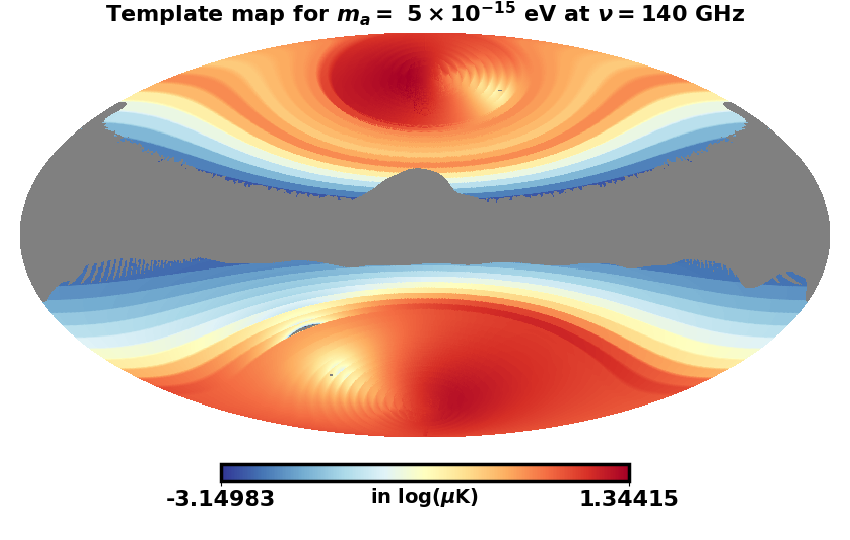}
         \caption{{Best-fit template map.}}
         \label{fig:mod_6}
     \end{subfigure}
\caption{ALP signal (smoothed to 51.1 arcmin) and best-fit template maps for $m_a = 5\times 10^{-15}$ eV at $\nu =$ 140 GHz ($g_{a\gamma} = 10^{-11} \, \mathrm{GeV^{-1}}$). }
\label{fig: sigmod_6}

\end{figure}

\begin{figure}[h!] 
\centering
\includegraphics[height=7cm,width=11cm]{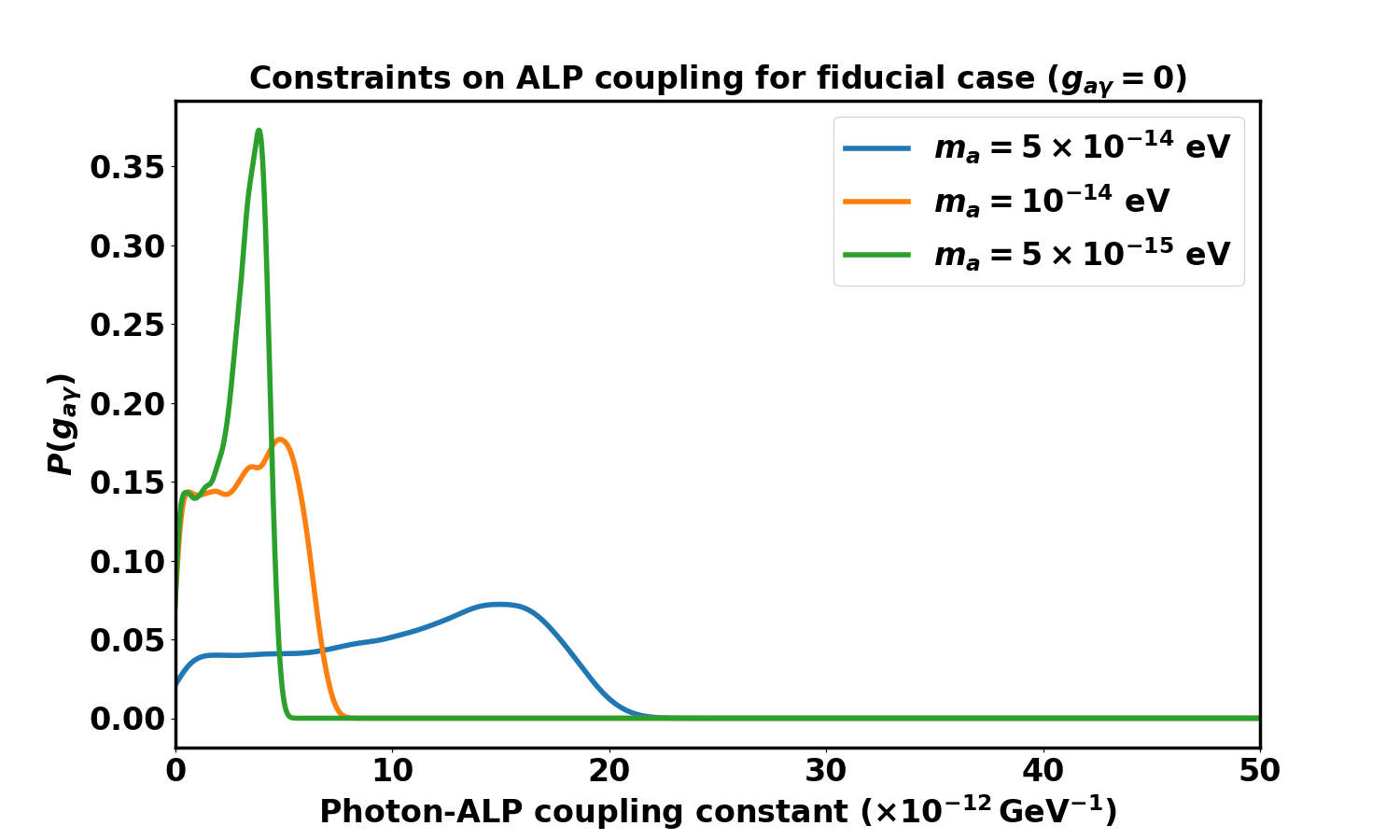}
\caption{Constraints on ALP coupling for fiducial case ($g_{a\gamma} = 0$) from different masses. The shift of the peak from zero indicates the presence of ILC residual noise due to the non-Gaussian nature of the foregrounds and due to the quality of fit in those patches. These can be accounted for using a template matching of these foregrounds, as well as using better models \cite{Mehta:2024pdz,Mehta:2024wfo}.}  
\label{fig:zero} 
\end{figure}

\begin{figure}[h!] 
\centering
\includegraphics[height=7cm,width=11cm]{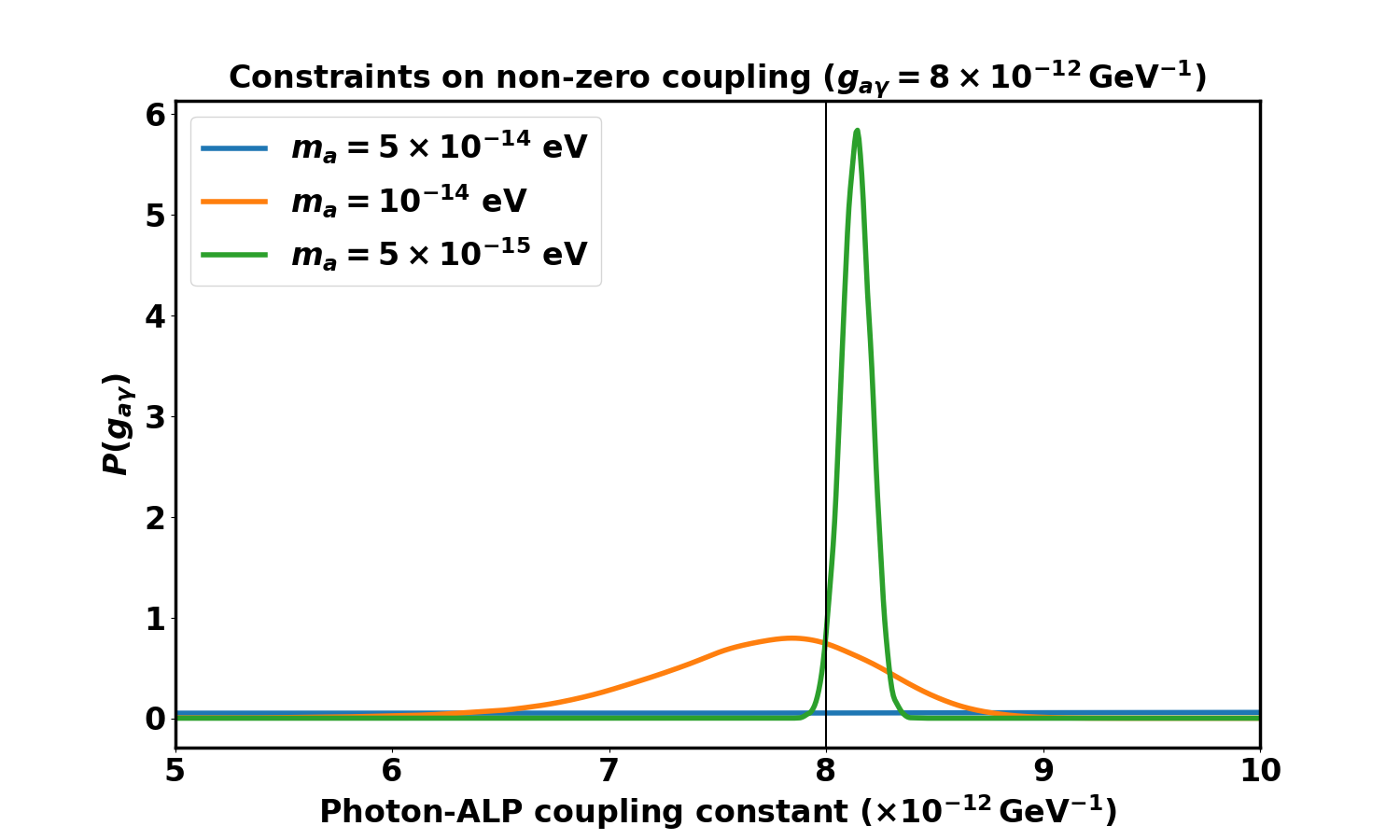}
\caption{Constraints on non-zero ALP coupling ($g_{a\gamma} = 8 \times 10^{-12} \, \mathrm{GeV}^{-1}$) from different masses.} 
\label{fig:nonzero} 
\end{figure}
Numerous models can be obtained by varying the parameter $d$ and functions $f(d,\theta)$ used in our modeling. We would now need to find the best-fit template, which fits the simulated signal very well, so that we can use it to get constraints on the ALP coupling. This requires us to perform a reduced $\chi_{red}^2$ (chi-squared) test, which can compare the efficiency of a model to fit the observed data. This is calculated as:
\begin{equation}
\chi_{red}^2 = \frac{1}{n-p}\sum_{i = 1}^{n} \frac{(S - M)^2}{\sigma^2} ,
    \label{eq:chisq}
\end{equation}
where $n$ refers to the number of pixels being fitted and $p$ is the number of parameters. The number of parameters estimated (which are 2 in our case: $A$ and $B$) for each latitudinal width ($\Delta \theta$) multiplied by the number of latitudinal widths for which the fitting was implemented. {The $\sigma$ refers to the pixel standard deviation of a different realization of the ILC-cleaned map, without ALP signal.}
We calculate the values for different templates
for the simulated maps of various masses, with the values of $d$ being varied from 12 to 20.5 kpc, in steps of 0.5 kpc, for all the model functions. {The reduced-$\chi^2$ values are calculated for various templates of masses $m_a = 5 \times 10^{-14}$ eV, $m_a = 10^{-14}$ eV, and $m_a = 5 \times 10^{-15}$ eV. In all three cases, the ALP coupling is taken to be $g_{a\gamma} = 10^{-11} \, \mathrm{GeV^{-1}}$.  The best-fit template is chosen as the one that is closer to the value of 1. We obtain the following $\chi_{red}^2$ values for the best-fit templates for various masses: $\chi_{red}^2 = 1.55$ (with Template-4) for $m_a = 5 \times 10^{-14}$ eV, $\chi_{red}^2 = 1.6$ (with Template-1) for $m_a = 10^{-14}$ eV, and $\chi_{red}^2 = 1.66$ (with Template-1) for $m_a = 5 \times 10^{-15}$ eV . A comparison of the best-fit templates from various models for the case of $m_a = 5\times 10^{-15}$ eV is shown in Appendix \ref{sec:compare}.  
The simulated polarization maps, along with their best-fit templates are shown in Fig. \ref{fig: sigmod_4} for $m_a = 5 \times 10^{-14}$ eV, in Fig. \ref{fig: sigmod_5} for $m_a = 10^{-14}$ eV, and in Fig. \ref{fig: sigmod_6} for $m_a = 5 \times 10^{-15}$ eV.} The signal maps are modeled at LiteBIRD's 140 GHz channel and  smoothed to the lowest resolution considered (51.1 arcmin). 

The best-fit template model $f(d,\theta)$ is expected to trace the effective resonant electron density at different latitudes. The fitting parameters will determine the strength and spatial shape of the signal. Thus, the $\chi_{red}^2$ values for the templates depend on the functions $f(d,\theta)$, the values $d$, and the fitting function parameters $A(\theta)$ and $B(\theta)$ at different $\theta$ values. The best-fit template will be determined by the combination of all these factors.  The reduced-$\chi^{2}$ estimation allows us to compare templates that may use a different number of parameters.  Better models can be constructed that can trace the profile variation in the galaxy and capture the small-scale features of the ALP signal, but our basic models can capture the large-scale features of the ALP signal. This template-based technique makes it possible to identify any spatial profile on the sky, such as the ALPs signal, but may miss out on the small-scale details. As a result, the template-based method will be particularly useful for fast search of ALP signals from microwave data, and for discovering any non-zero signal. If there are non-zero signals present, then one can perform detailed modeling of the 3-D ALPs signal using the 3-D model of electron density and magnetic field, and also capture the small-scale features to improve both accuracy and precision. Though this model can be more accurate, but won't be possible for a search for the ALPs signal in a Bayesian way, due to the computational cost of making the full sky maps.

\subsection{Estimating the photon-ALPs coupling strength from the Milky Way}
\label{sec:coupling}
The best-fit template can then be used to get constraints on the photon-ALP coupling $g_{a\gamma}$. This can be done using temperature as well as polarization observations. We perform a joint Bayesian inference of the coupling constant from temperature and polarization maps using the Markov Chain Monte Carlo (MCMC) algorithm. This uses the Bayes theorem \cite{trotta2017bayesian,trotta2008bayes,heavens2011bayesian,hobson2010bayesian}, which goes as:
\begin{equation}
P(g_{a\gamma}|S_{\rm{ilc}})\propto \mathcal{L}(S_{\rm{ilc}}|g_{a\gamma})\pi(g_{a\gamma}),
    \label{eq:posterior}
\end{equation}
where the posterior  ($P$) on the parameter $g_{a\gamma}$, given the mock data map goes proportional to the prior ($\pi$) on the coupling $g_{a\gamma}$ and likelihood ($\mathcal{L}$) of the mock data map given the ALP coupling. We use emcee \cite{Foreman_Mackey_2013} to perform Monte Carlo Markov Chain (MCMC) parameter estimation. We use a uniform prior on the ALP coupling $g_{a\gamma}$ between a range of 0 to $6.6 \times 10^{-11} \, \mathrm{GeV^{-1}}$ (current bounds from the CERN Axion Solar Telescope (CAST) \cite{2017}). { Since we are using the sum of polarized intensity squared values at various pixels, the distribution of such values for multiple realizations will have a non-zero mean. We create 500 different map realizations ($R$) to calculate the mean and variance of the distribution of these non-zero values. We use the Gaussian likelihood that goes as:
\begin{equation}
\log \mathcal{L} = - \frac{ \sum_{i = 1}^{n} (S^{i})^2 - (g_{a\gamma} / g*)^4\sum_{i = 1}^{n} (M^i)^2 - \langle R \rangle}{2 \,  \sigma_R ^2} - 0.5 \log[2\pi \,  \sigma_R^2 ],
\label{eq:axlike}
\end{equation}
where $n$ refers to the number of pixels, $S$ is the ILC-cleaned polarized intensity map, and $M$ refers to the ILC-cleaned best-fit polarized intensity template map. For our analysis, the templates are fitted at the standard coupling constant value of $g* = 10^{-11} \, \mathrm{GeV}^{-1}$. Here $ R $ is calculated from multiple realizations of ILC-cleaned fiducial maps (with ALP coupling $g_{a\gamma} = 0$) with components CMB, foregrounds, and instrument noise:
 \begin{equation}
  R  =  \sum_{i=1}^{n} (D^{i})^2  , 
 \end{equation}
 }
where $D$ refers to the various ILC-cleaned fiducial map realizations (with zero ALP coupling).  
The quantity $\sigma _R$ corresponds to the standard deviation of the distribution of $R$ values from various realizations. This takes into account not only the correlation for a component at multiple frequencies, but also the correlations between different components of the maps (such as synchrotron and dust). This is required to get an unbiased result on the estimation of a new signal in comparison to a fiducial cosmological model with no ALPs.

We show our results for the different ALP masses in Fig. \ref{fig:zero} for the fiducial case of non-existence of ALPs ($g_{a\gamma} = 0$). The best constraints are obtained for the case of $m_a = 5 \times 10^{-15}$ eV at $g_{a\gamma} < 4.5 \times 10^{-12} \, \mathrm{GeV}^{-1}$ at 95\% C.I. For the cases of $10^{-14}$ and $5\times 10^{-14}$ eV, we are able to obtain bounds of $g_{a\gamma} < 6.5 \times 10^{-12} \, \mathrm{GeV}^{-1}$ and $g_{a\gamma} < 1.9 \times 10^{-11} \, \mathrm{GeV}^{-1}$ respectively at 95 \% C.I. This is because the ALP distortion signal is stronger for the case of low-mass ALPs. The constraints are about an order of magnitude better than the current bounds from  CAST \cite{2017}. The shift of the peaks from zero depends on the quality of fit and also poorly modeled spatial variation of the galactic foregrounds. The former can be improved by better modeling using complex models that capture various signal features, or a 3-D modeling of various patches, if possible, with an accurate galactic electron density and magnetic field model. The latter effect can be mitigated by using cleaning techniques such as the template matching of foregrounds \cite{Mehta:2024pdz,Mehta:2024wfo}. We show how to understand the underlying possible systematic uncertainty in the inference of photon-ALPs coupling strength in Sec. \ref{sec:partial}.   

The constraints for the case of non-zero coupling are shown in Fig. \ref{fig:nonzero}, where we have taken $g_{a\gamma} =  10^{-11}$ GeV$^{-1}$. The cases of $m_a = 10^{-14}$ eV and $m_a = 5 \times 10^{-15}$ eV are able to constrain the coupling with lower as well as upper limits. {This shows the robustness of the sky-template-based technique to be able to constrain non-zero couplings as well. The quality of the template fitting will decide the quality of the constraints and the shift from the true value. This, in turn, depends on the ability of the template to be able to capture both the large and small-scale features of the sky signal. The left-right asymmetry in the signal map is an important feature that will have to be captured if better modeling of the signal is needed. Here again, the constraints are affected by the quality of template fit as well as the spatial distribution of the galactic foregrounds, and could be better dealt with if template matching of foregrounds is used. If the foregrounds are not modeled well, it can introduce a bias in the inference of the coupling strength. These systematics can be better explored by using a partial sky analysis, which we describe in the next section (Sec. \ref{sec:partial}).}  


\subsection{Estimation of systematic uncertainty using position-dependent inference of the photon-ALPs coupling}
\label{sec:partial}

In the previous section, we estimated the full-sky averaged constraints on the ALPs signal and showed the performance of the sky-template-based technique. However, in order to develop a robust detection technique, we need to understand the impact of mis-modeling of the non-isotropic galactic foregrounds and ALPs signal template on the inference of the photon-ALPs coupling constraints, and how it can be mitigated in the analysis. This is particularly important because the signal of ALPs which we are trying to measure is a map-level inference (using a spatial template). As a result, any map-level contamination which are not well modeled, can be a potential source of contamination, and it is important to show the vulnerability of the inference of the photon-ALPs coupling strength due to these kinds of problems.    

\begin{figure}[h!] 
\centering
\includegraphics[height=8cm,width=14cm]{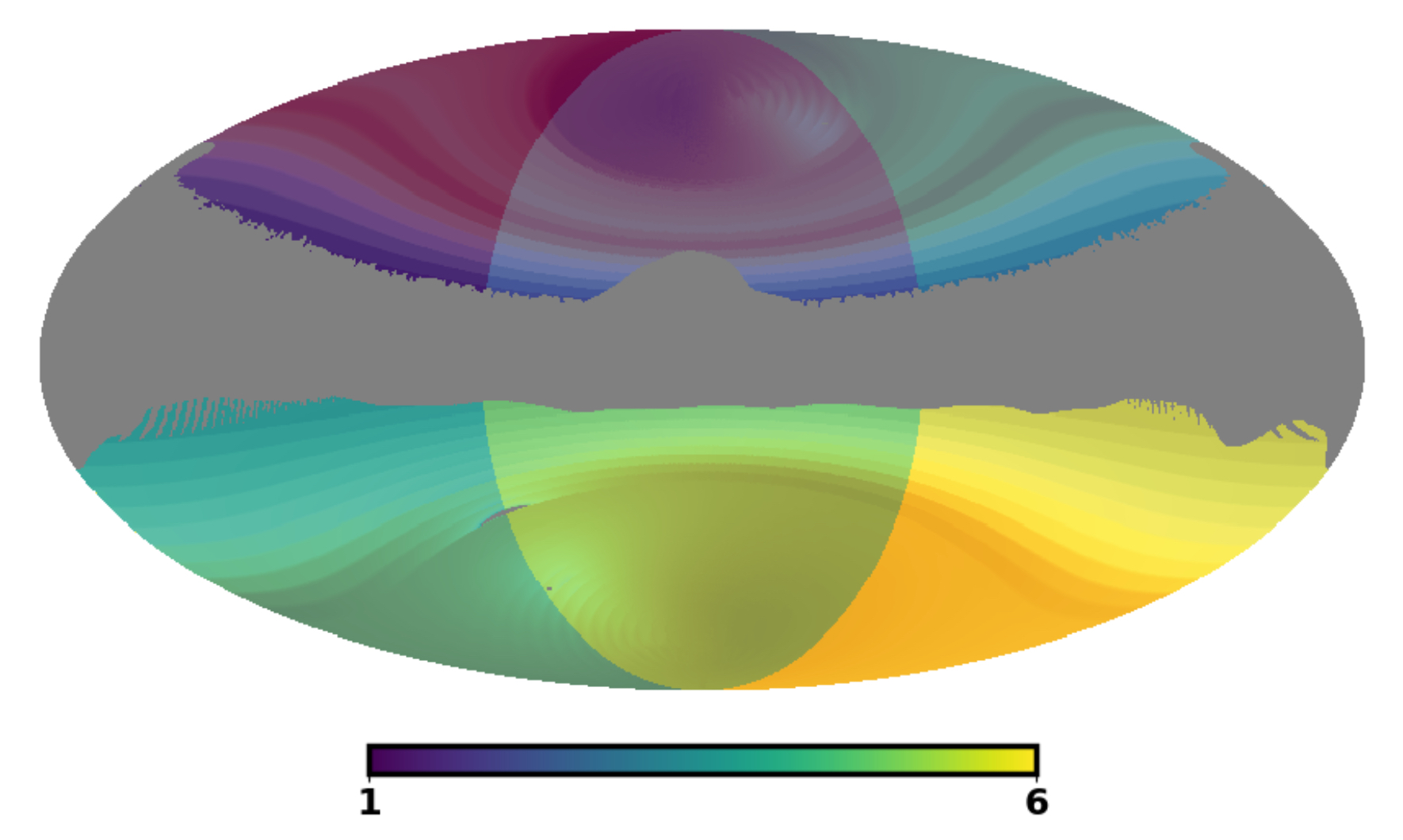}
\caption{Patch numbers for different sky portions. This is superposed on the best-fit template map for ALP mass $m_a = 5\times 10^{-15}$ eV to indicate the separate features which are captured by different patches. This enables us to make independent inferences from different portions of the sky, depending on the template fit in different sky regions.}
\label{fig:patch} 
\end{figure}

We show that direction-dependent inference of the photon-ALPs coupling strength by dividing the sky into nearly equal sky areas (as shown in Fig. \ref{fig:patch}) can shed light on the presence of possible systematics. The sky direction-dependent analysis can exhibit a sky patch-dependent inference of the coupling strength due to poorly modeled foregrounds and signal templates, arising from our ignorance of the statistical properties of galactic foregrounds, galactic electron density, and galactic magnetic field.

\begin{figure}[h!] 
\centering
\includegraphics[height=7cm,width=12cm]{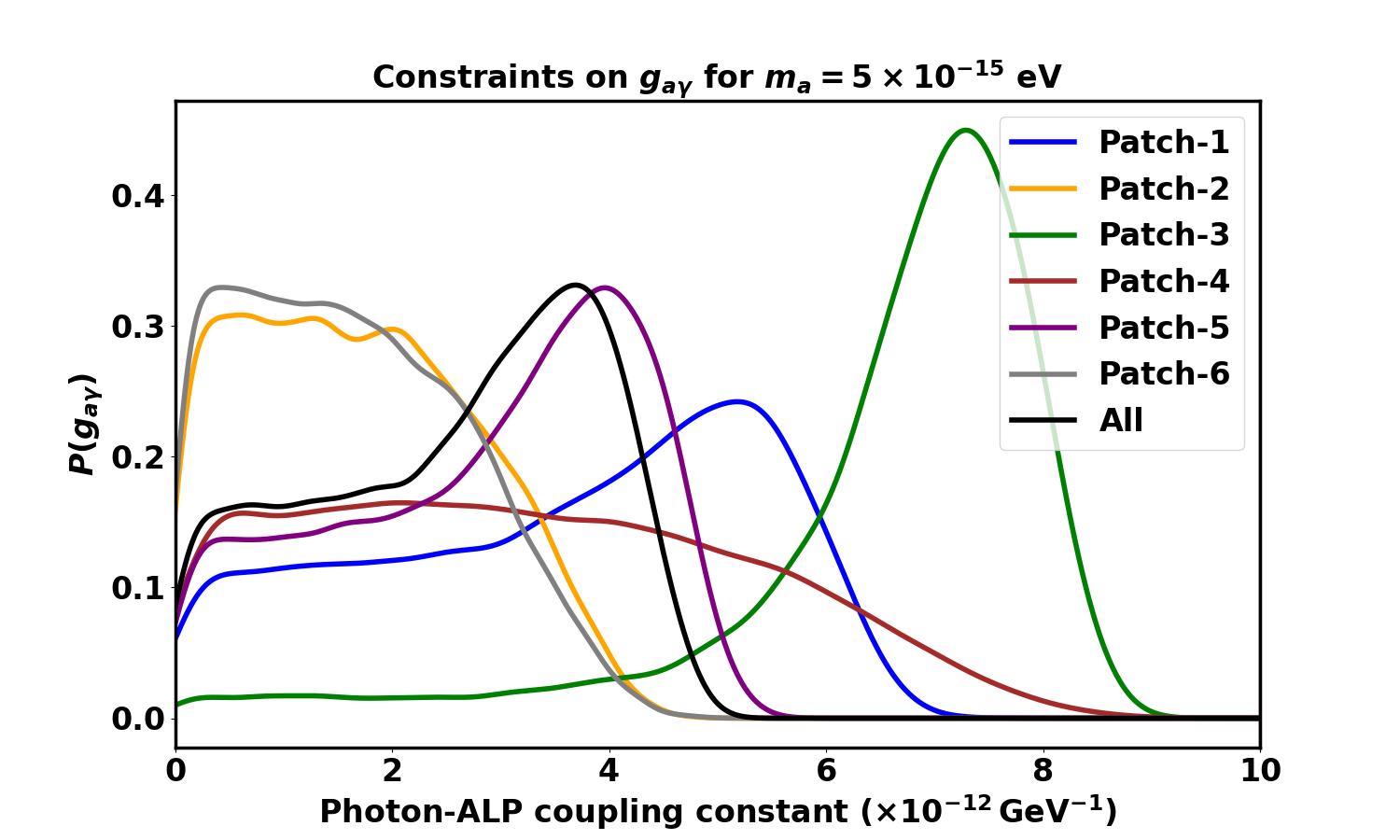}
\caption{Constraints on zero coupling from different patches.} 
\label{fig:multiple0} 
\end{figure}

\begin{figure}[h!] 
\centering
\includegraphics[height=7cm,width=12cm]{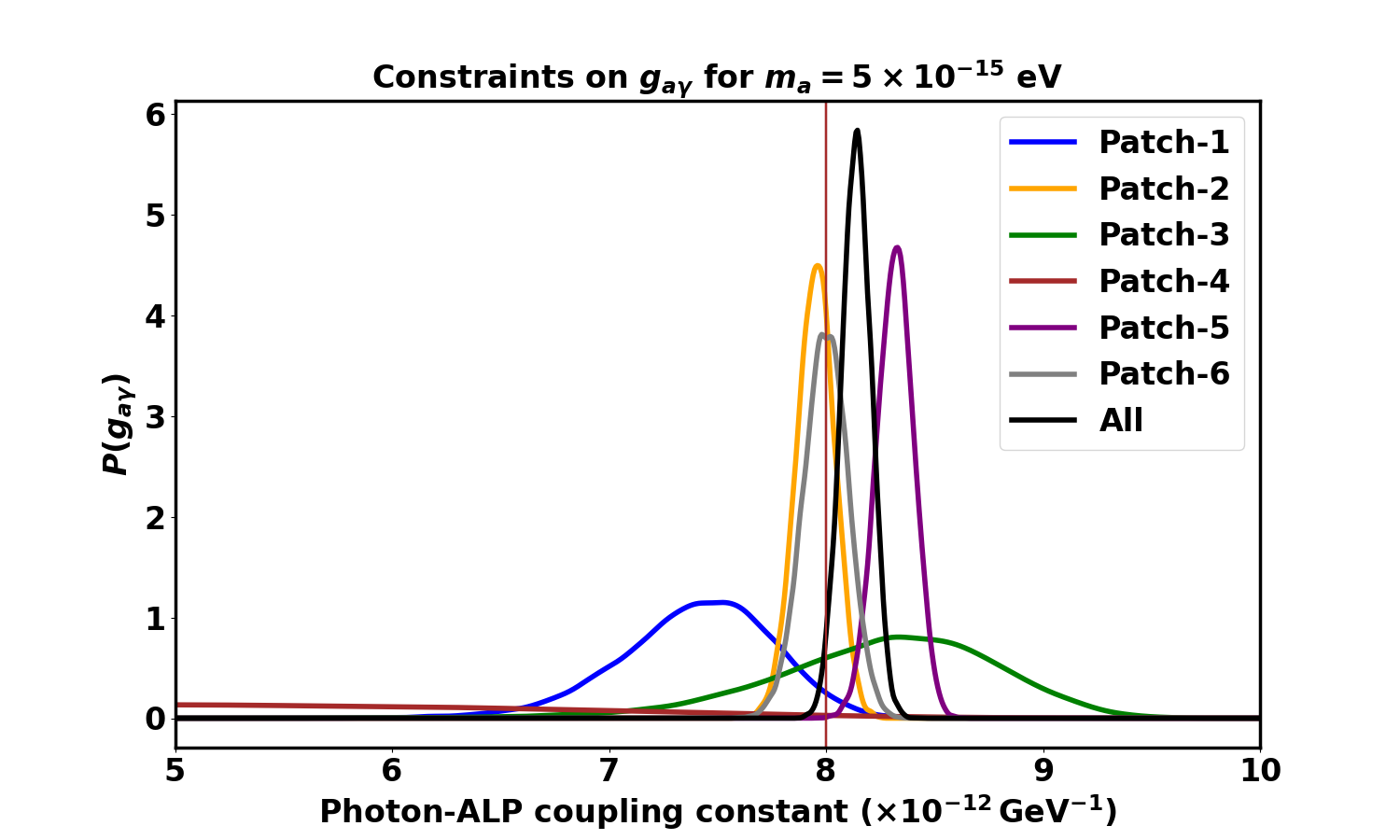}
\caption{Constraints on non-zero coupling ($g_{a\gamma} = 8 \times 10^{-12} \, \mathrm{GeV}^{-1}$) from different patches.} 
\label{fig:multiple8} 
\end{figure}


{We analyze the mock ALPs signal by obtaining constraints from different regions of the sky as shown in Fig. \ref{fig:patch}. The constraints from different regions for ALP mass $m_a = 5 \times 10^{-15}$ eV is shown in Fig. \ref{fig:multiple0} and Fig. \ref{fig:multiple8} for fiducial case ($g_{a\gamma} = 0$) and ALP coupling $g_{a\gamma} = 8 \times 10^{-12} \, \mathrm{GeV}^{-1}$ respectively. As shown in Fig. \ref{fig:multiple0},  the sky direction dependent constraints exhibit variation from  $g_{a\gamma} = 0$ up to a factor of 2-3 variation in the constraints on the coupling strength from different sky patches. This result shows important aspects: (i) the direction-dependent inference of coupling strength can directly capture the patches, which shows an outlier behavior, and (ii) the combined measurement from all the patches is well consistent with the injected value (as shown in Fig. \ref{fig:zero}). The constraints from different sky patches depend on the possible strength of the signal in those patches of the sky, the quality of the template-fit in those patches, and the impact of non statistically isotropic galactic foregrounds.  

{For the non-zero coupling case, the injected coupling ($g_{a\gamma} = 8 \times 10^{-12} \, \mathrm{GeV}^{-1}$) is inferred for various sky patches, showing about a factor of two variation between different sky patches. The constraints from patches 2 and 6 look most reliable, while the ones obtained from patches 4 and 5 are strongly impacted. The result for the non-zero photon-ALPs coupling case also shows that, though there is a sky position dependence in the inference of the signal due to limitations in the modeling of the galactic foregrounds, galactic electron density, and the magnetic field, the all-sky inference of the signal agrees with the injected value. One of the main contaminations to the ALP signal comes from the non-Gaussian thermal dust emission, which contributes significantly at frequencies greater than 70 GHz. Accounting for this non-Gaussianity of foregrounds (especially, dust) will significantly improve the constraints. Thus, a better modeling incorporating both the ALP signal and foreground contaminations could improve the measurements.}} 

In summary, our analysis shows that although the photon-ALPs coupling strength inference from Milky Way is limited by accuracy in the modeling of galactic foregrounds, galactic electron density, and galactic magnetic field, one can perform a fast sky-template-based analysis to search for any non-Gaussian sky signal on the sky and can understand the possible systematics associated with the inference of the photon-ALPs coupling strength by performing a sky direction dependent analysis. Such analysis will not help in identifying the outlier patches, but will also help in estimating the underlying systematic uncertainty in the measurement of the photon-ALPs coupling strength along with the statistical uncertainty. Even though in the near future, an accurate 3-D model of galactic electron density and magnetic field will be difficult, a robust analysis pipeline will be able to infer the presence of ALP distortion in the sky in a robust way, by capturing both statistical and systematic uncertainties.


\section{Conclusion}
The ALP distortion signal from our galaxy due to CMB photon-ALP resonant conversion depends on the electron density and magnetic field profiles in our galaxy.
The galactic halo and disk consist of ionized gases and free-electron densities. The magnetic fields in this ionized plasma are of the order of $\mu$G and determine the plasma hydrodynamics \cite{jansson2012galactic,jansson2012new,oppermann2012improved,adam2016planck}. The electron densities and magnetic field profiles in our galaxy are characterized by small-scale turbulence. This makes it difficult to make precise measurements of these profiles. The ALP signal shows non-Gaussian features \cite{mehta2025turbulence}, so it is difficult to probe it using only the power spectrum information of the CMB. In addition, simulating the ALP signal for these profiles is computationally heavy and calls for the need for faster methods to search for the ALP distortion signal. 

In this work, we have developed a sky-template-based approach to search for the galactic ALP distortion signal. This method captures the sky-shape of the ALP signal based on the large-scale electron densities and magnetic fields known to us. A number of templates can be used to search for the ALP signal in a much faster approach. This approach also captures the non-Gaussian aspect of the ALP signal and does not require polarization information, as it works at the intensity level (see Sec. \ref{sec:ALP_template}). We use the ILC weights to clean the ALP signal using a quadratic combination of the multi-frequency maps. We show for a few models how this method can be used to find the best-fit templates for ALPs of masses $5 \times 10^{-14}$ eV, $ 10^{-14}$ eV and $5 \times 10^{-15}$ eV by estimating the reduced-$\chi^2$ value for the various templates and selecting the one closest to the value of 1 (see Sec.  \ref{sec:bestfit}). The best-fit template is then used to constrain the ALP coupling for the fiducial case ($g_{a\gamma} = 0$) and for the case of non-zero coupling ($g_{a\gamma} = 10^{-11} \, \mathrm{GeV}^{-1}$) as shown in Sec. \ref{sec:coupling}. These constraints show that our sky-template-based technique is efficient enough in being able to search for ALPs. For the fiducial case, the constraints for the various masses at 95\% C.I. are obtained as: $g_{a\gamma} < 4.5 \times 10^{-12} \, \mathrm{GeV}^{-1}$ for $m_a = 5 \times 10^{-15}$ eV, $g_{a\gamma} < 6.5 \times 10^{-12} \, \mathrm{GeV}^{-1}$ for $m_a =  10^{-14}$ eV and $g_{a\gamma} < 1.9 \times 10^{-11} \, \mathrm{GeV}^{-1}$ for $m_a = 5 \times 10^{-14}$ eV. {These bounds are over an order of magnitude better than the current bounds from CAST ($g_{a\gamma} < 6.6 \times 10^{-11} \, \mathrm{GeV}^{-1}$) \cite{2017}. Using CMB observations of galaxy clusters with high-resolution surveys, such as the Simons Observatory (SO) \cite{Ade_2019}  and CMB-S4 \cite{abazajian2016cmbs4}, we will be able to obtain bounds on ALP coupling of up to $g_{a\gamma} < 5.2 \times 10^{-12} \, \mathrm{GeV}^{-1}$ and $g_{a\gamma} < 3.6 \times 10^{-12} \, \mathrm{GeV}^{-1}$ respectively for ALP mass $m_a = 10^{-13}$ eV \cite{Mehta:2024wfo}. Using these surveys to probe the diffuse ALP background from unresolved clusters for ALPs of masses $10^{-15}$ to $10^{-11}$ eV, we can obtain bounds of $g_{a\gamma} < 10^{-12} \, \mathrm{GeV}^{-1}$ \cite{Mehta:2024pdz}. Combining these observations with those of Milky Way observations by LiteBIRD will make it possible to place much stronger constraints on the ALP coupling. Using future surveys such as CMB-HD \cite{sehgal2019cmbhd}, or cleaning techniques such as the template matching of foregrounds \cite{Mehta:2024pdz,Mehta:2024wfo} will provide even stronger bounds on the coupling constant.} 
Finally, we show the inference of the coupling from different patches of the sky in Sec. \ref{sec:partial}, showing how one can estimate and mitigate systematic uncertainties due to poorly modeled galactic foregrounds, electron density, and magnetic field.

{This novel technique will enable searching for galactic ALPs distortion signals from data from wide-field surveys, such as LiteBIRD. LiteBIRD will enable efficient foreground cleaning using its multiple frequency channels and almost a full-sky field of view.} The template search technique is a robust and fast technique that can also be used to search for other non-Gaussian signals, for which the power spectrum is not a sufficient measure to probe the signal. It captures the large-scale aspects of the signals and can be used to make cosmological inferences using them. This technique ensures a computationally inexpensive search for the galactic ALPs signal for future wide-field CMB survey LiteBIRD to illuminate on the photon-ALPs coupling strength over an unexplored parameter space.

\label{sec:conclude}

\appendix
\section{Template Model comparison }
\label{sec:compare}
\begin{figure}[h!]
     \centering
     \begin{subfigure}[h]{0.45\textwidth}
         \centering    
\includegraphics[height=5cm,width=7.3cm]
         {images/sigmap_512_6.png}
         \caption{{Galactic ALP signal map.}}
         \label{fig:sig5_6}
     \end{subfigure} 
\\
     \centering
     \begin{subfigure}[h]{0.45\textwidth}
         \centering    
\includegraphics[height=5cm,width=7.3cm]
         {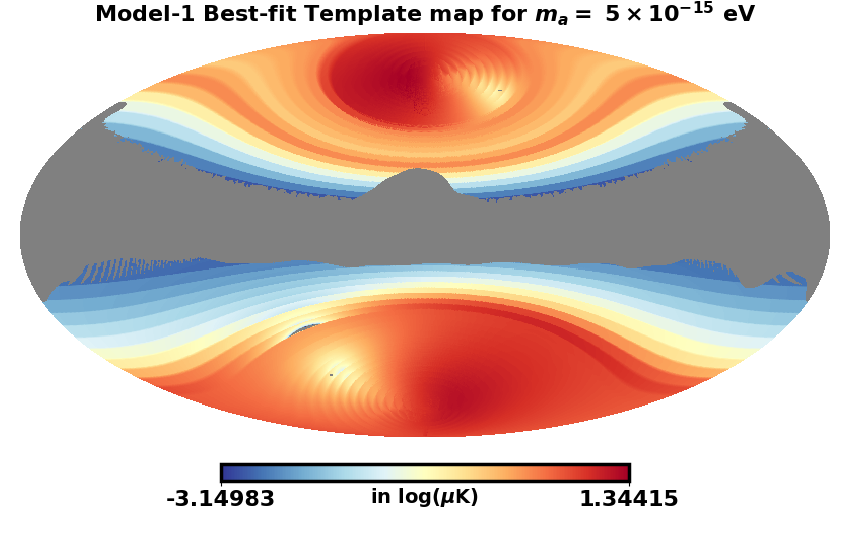}
         \caption{{Overall Best-fit Template from Template-1}}
         \label{fig:mod1_6}
     \end{subfigure} 
     \hspace{0.01cm} 
     \begin{subfigure}[h]{0.45\textwidth}
         \centering
    \includegraphics[height=5cm,width=7.3cm]
         {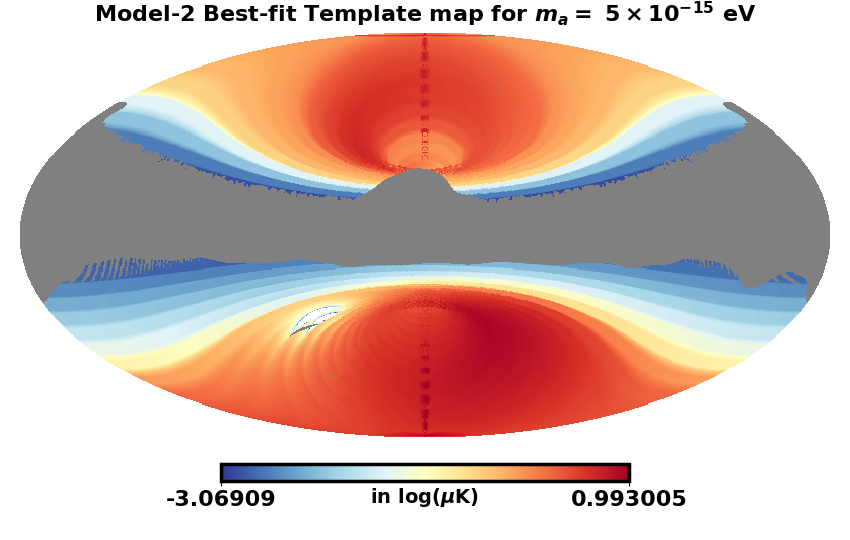}
         \caption{{Template-2 Best-fit Template}}
         \label{fig:mod2_6}
     \end{subfigure}

     \centering
     \begin{subfigure}[h]{0.45\textwidth}
         \centering    
\includegraphics[height=5cm,width=7.3cm]
         {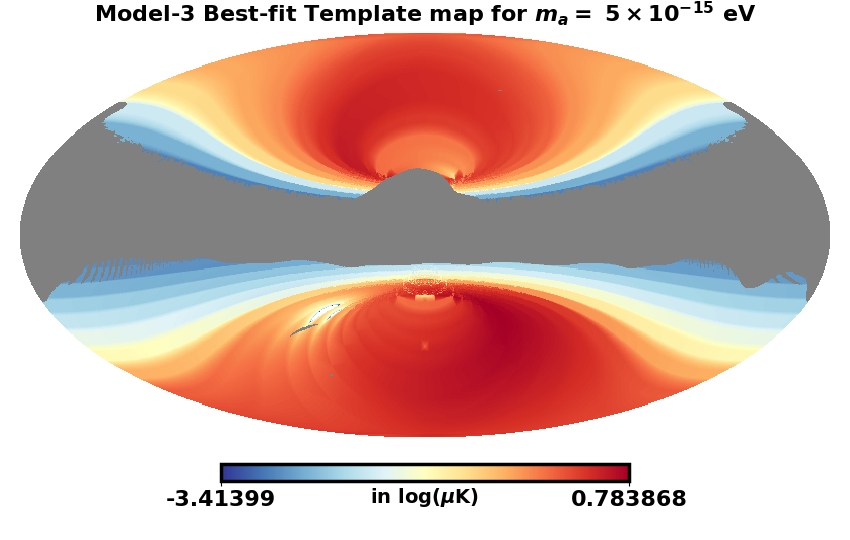}
         \caption{{Template-3 Best-fit Template}}
         \label{fig:mod3_6}
     \end{subfigure} 
     \hspace{0.01cm} 
     \begin{subfigure}[h]{0.45\textwidth}
         \centering
    \includegraphics[height=5cm,width=7.3cm]
         {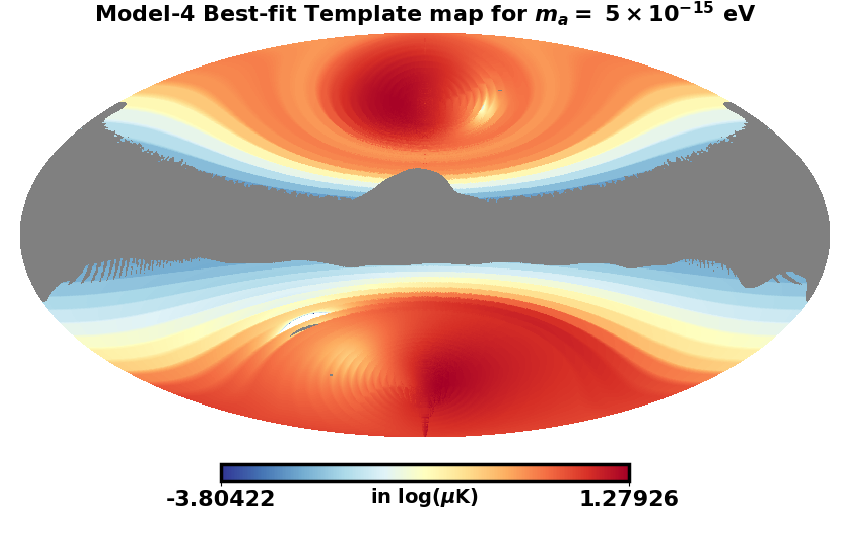}
         \caption{{Template-4 Best-fit Template}}
         \label{fig:mod4_6}
     \end{subfigure}
\\
\caption{Best-fit template maps for $m_a = 5\times 10^{-15}$ eV at $\nu =$ 140 GHz ($g_{a\gamma} = 10^{-11} \, \mathrm{GeV^{-1}}$). }
\label{fig: allmod}
\end{figure}

Here we show the comparison of the best-fit template maps for various models for the case of $m_a = 5 \times 10^{-15}$ eV and $g_{a\gamma} = 10^{-11} \, \mathrm{GeV}^{-1}$. The $\chi_{red}^2$ values for the best-fit template maps (shown in Fig. \ref{fig: allmod}) from various models are: 1.66 for Template-1; 2.39 for Template-2; 2.64 for Template-3; 1.78 for Template-4. The best-fit template for Template-1 does indeed capture the features of the ALP signal on large scales. This shows that our sky-template-based technique can be improved by using better models that can capture the small-scale features of the ALP signal, such as the right-left asymmetry, which leads to varying constraints from different sky patches.

\acknowledgments
    This work is a part of the $\langle \texttt{data|theory}\rangle$ \texttt{Universe-Lab}, supported by the TIFR  and the Department of Atomic Energy, Government of India. The authors express their gratitude to the $\langle \texttt{data|theory}\rangle$ \texttt{Universe-Lab} and the TIFR CCHPC facility for meeting the computational needs. 
 Also, the following packages were used for this work: Astropy \cite{astropy:2013,astropy:2022,astropy:2018},
, NumPy \cite{harris2020array}
CAMB \cite{2011ascl.soft02026L}, SciPy \cite{2020SciPy-NMeth}, SymPy \cite{10.7717/peerj-cs.103}, Matplotlib \cite{Hunter:2007}, emcee \cite{Foreman_Mackey_2013}, HEALPix (Hierarchical Equal Area isoLatitude Pixelation of a sphere)\footnote{Link to the HEALPix website http://healpix.sf.net}\cite{2005ApJ...622..759G,Zonca2019} and PySM \cite{Thorne_2017}.

\bibliography{references.bib}








\end{document}